\documentclass[%
 aip,
 amsmath,amssymb,
 reprint,%
]{revtex4-1}

\usepackage{graphicx}
\usepackage{dcolumn}
\usepackage{bm}

\usepackage[utf8]{inputenc}
\usepackage[T1]{fontenc}
\usepackage{mathptmx}
\usepackage{etoolbox}
\usepackage{xcolor}

\makeatletter
\def\@email#1#2{%
 \endgroup
 \patchcmd{\titleblock@produce}
  {\frontmatter@RRAPformat}
  {\frontmatter@RRAPformat{\produce@RRAP{*#1\href{mailto:#2}{#2}}}\frontmatter@RRAPformat}
  {}{}
}%
\makeatother
\begin{document}

\preprint{AIP/123-QED}

\title{Viscoelasticity of biomimetic scale beams from  trapped complex fluids}

\author{Pranta Rahman Sarkar}
\affiliation{Mechanical and Aerospace Engineering, University of Central Florida, Orlando}

\author{Outi Tammisola}
\affiliation{Mechanical Engineering, KTH-Royal Institute of Technology, Stockholm}

\author{Ranajay Ghosh}
\affiliation{Mechanical and Aerospace Engineering, University of Central Florida, Orlando}
\email{ranajay.ghosh@ucf.edu}

\date{\today}

\begin{abstract}


We investigate the nonlinear viscoelastic behavior of a biomimetic scale-covered beam in which shear-dependent complex fluids are trapped between overlapping scales under bending loads. These fluids mimic biological mucus and slime layers commonly enveloping the skins found in nature. An energy-based analytical model is developed to quantify the interplay between substrate elasticity, scale geometry, and fluid rheology at multiple length scales. Constant strain rate and oscillatory bending are examined for Newtonian, shear-thinning, and shear-thickening fluids. The analysis reveals unique, geometry- and rate-dependent viscoelastic response, distinct from classical mechanisms such as material dissipation, frictional resistance, or air drag. Energy dissipation is shown to emerge from a nonlinear coupling of tribological parameters, fluid rheology, and system kinematics, exhibiting distinct regime-differentiated characteristics. The model captures the competitions and cooperations between elastic and geometrical parameters to influence the viscoelastic behavior and lead to geometry and rheology scaling laws for relative energy dissipation. The pronounced nonlinearity in the moment–curvature relationships, along with the geometry-controlled regimes of performance, highlights the potential for using tailored and engineered complex inks for soft robotics and smart damping systems.

\end{abstract}

\maketitle


\section{Introduction}

\begin{figure*} [t]
\begin{center}
\includegraphics[scale = 0.5]{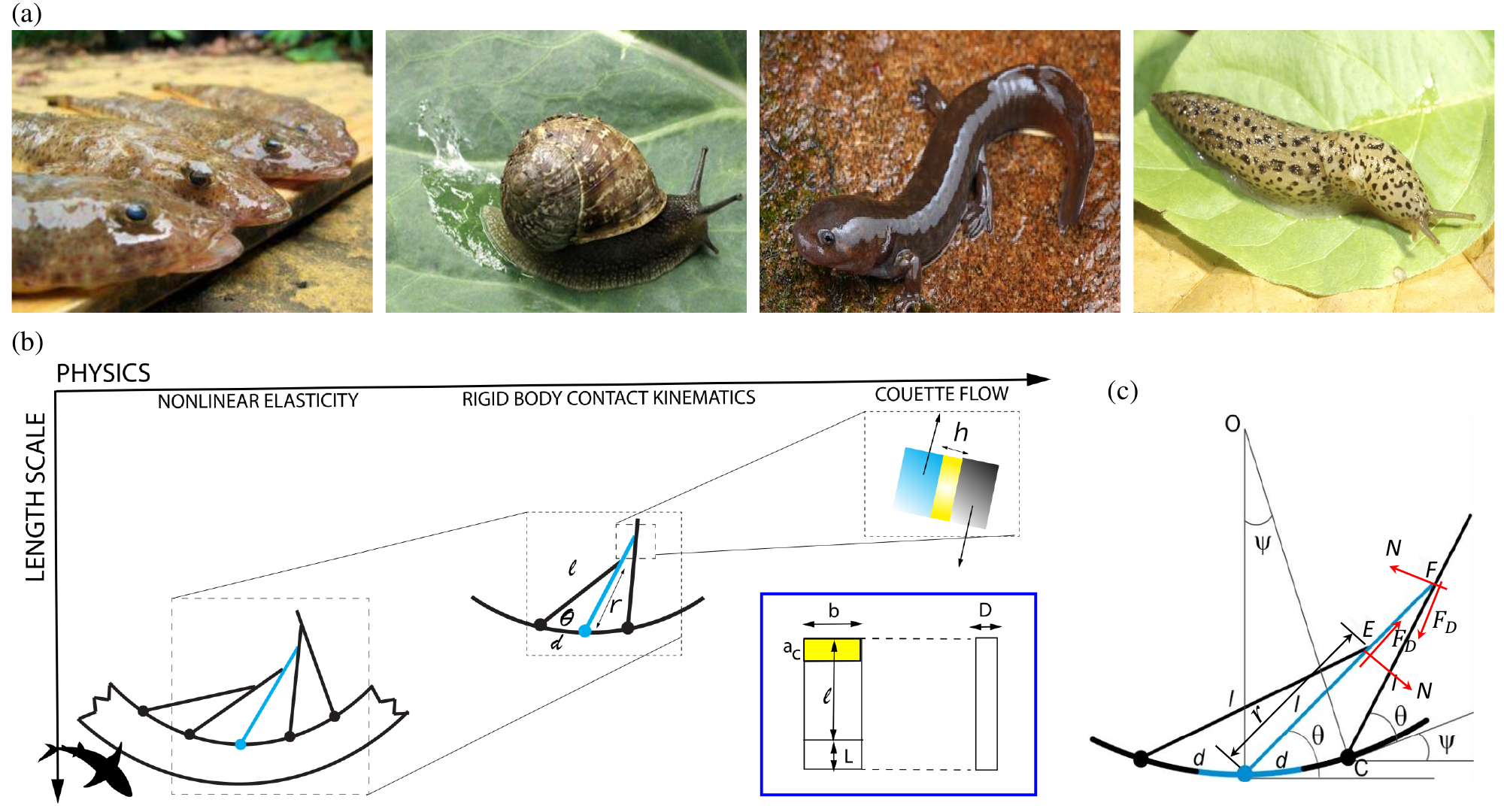}
\end{center}
\caption{(a) Natural examples of slime-covered organisms, including fish (image reproduced with permission from © Reefs.com \cite{Reefs2020}), snail (© Sarah Swanson \cite{Swanson2020}), salamander (© Jake Scott, iNaturalist \cite{Scott2024}), and slug (© David Shetlar, The Ohio State University Extension \cite{OSUExtension2024}), demonstrating the role of biological slime in protection, lubrication, and load-bearing functionality. (b) Schematic illustration of the multiscale-multiphysics nature of the problem - from the structure level to RVE (Representative Volume Element) to the lubrication length scale from left to right. (c) Detailed geometric relationships showing scale articulation, sliding kinematics, normal and dissipative contact forces under curvature-induced deformation (Figure 1(c) is adopted from \cite{ghosh2017non} with slight modifications).}
\label{6-Fig1}
\end{figure*}

Fish scales provide mechanical protection \cite{lovegrove2001evolution, bruet2008materials, yang2013natural}, aid in camouflaging \cite{gower2003scale, prum2006anatomically, doucet2009iridescence, denton1970review}, and contribute to the hydrodynamic efficiency \cite{drucker2002experimental, borazjani2008numerical, oeffner2012hydrodynamic} of swimming. These natural adaptations have evolved over millions of years, resulting in sophisticated designs that balance protection, locomotion, flexibility, and functions such as optics \cite{chintapalli2014fabrication, martini2017comparative, martini2016stretch,ibanez2009variation, rudykh2015flexibility}. Although external fluid interaction with the scales has attracted intense attention, another aspect of fluid-structure interaction has gone unnoticed so far; the role of fluid trapped between the scales in the form of slimes, which are commonly found on a variety of fishes \cite{fudge2005composition, wainwright2024hydrodynamic, fischer2025slippery, wainwright2017mucus, chaudhary2019unravelling}. Such slimy secretions are not unique to fishes and are also observed serving functional roles in snails, salamanders, and slugs \cite{rashad2023biological, barajas2023probing} (Fig.~\ref{6-Fig1}(a)). When trapped between the scales, these slimes can aid in locomotion, offer protection, and reduce gripping potential, thereby providing a significant boost to survivability \cite{yan2022slime, wainwright2017mucus}. Typically, these slimes behave as complex fluids \cite{siddiqui2001undulating, ali2016bacterial, mahomed2007gliding, ewoldt2022designing, bowers2023modeling}, tending toward shear-thinning behavior. The trapped fluid can significantly influence the dynamic response of the scaled system, which in turn affects overall locomotion and dynamic efficiency \cite{chaudhary2019unravelling, yan2022slime, mandel2020understanding, wang2020self}. While slimes are known to directly help in the locomotion of snails and other gastropods \cite{rashad2023biological, barajas2023probing}, their effect on the dynamics of fish scale like systems is less well understood. So far the literature has put more focus on external hydrodynamics \cite{drucker2002experimental, borazjani2008numerical, oeffner2012hydrodynamic}, neglecting internal trapped fluid effects. We find in this study that trapped fluid plays a critical role in the mechanical behavior of these substrates. This work thus opens the fascinating possibility of specially designed bio-inspired complex fluids (\textit{robot slime}) that can be another important \textquoteleft structural\textquoteright \space element of the system; such slimes can be either shear thinning, Newtonian, or even shear thickening as per the purpose of the robotic system. Understanding this physics thus opens up a new avenue of research in soft robotics and smart materials \cite{sadati2015stiffness, sire2009origin}. 

From a material standpoint, the trapped fluid directly contributes to rate effects and dissipation in the overall structure during substrate deformation, giving it a viscoelastic character. However, the viscoelasticity here is fundamentally different from traditional material sources, depending upon the fluid dynamics and the kinematics of the system. Further complexity in the viscoelastic response arises due to the relatively early onset of nonlinear elasticity (even in small strains) as the scales begin to engage and start sliding over each other. The collective sliding kinematics introduces nonlinear strain stiffening even when friction is absent \cite{ghosh2014contact, ebrahimi2019tailorable}. This sliding behavior gives rise to at least three broad regimes of elastic behavior of dry systems - linear (before engagement), nonlinear strain stiffening (post engagement) and finally quasi-rigid locked states where further sliding is not possible without substantially bending the much stiffer scales \cite{ghosh2014contact, ebrahimi2019tailorable, ghosh2016frictional, ebrahimi2020coulomb}. These observations raise important questions about the analogous role of scales in causing viscoelasticity. For instance, the scales sliding leading to a rise in viscous forces simultaneously with nonlinear elasticity leading to competing behaviors. These fascinating landscape of nonlinear viscoelasticity is still relatively understudied either computationally or experimentally. The large number of contacts, rheology of complex fluids, and imposing their coupling make direct numerical simulation of both structural and fluid problem a formidable challenge, too demanding for most commercial software even for relatively simple load cases \cite{tatari2023bending}. Similarly, hydrodynamic or tribological experiments that precisely target the coupled phenomena has proven to be difficult due to the overlapping and changing geometry, opacity of scales, and difficulty in measuring and observing the behavior of complex fluids. These challenges, are in addition to the the already compounding combinatorial complexity of the parameters, prevent simulation- or experiment-driven physical understanding of these systems.

Under these circumstances, analytical models play a critical role in advancing fundamental understanding of such systems. They serve as essential tools to guide simulations, material selections, and experimental campaigns by providing sample geometries and loads of interest. They also provide the validation and verification sets and cases to help complete the validation loop. More importantly, these models help to isolate the effects of various physical parameters and guide simulations and experimental campaigns. Although analytical models are based on many simplifications, previous studies have shown that even when the restrictions of analytical models are removed, the divergences are still broadly in line with the trends of the simplified models \cite{ghosh2017non, ali2019bending}. Therefore considering the significance of analytical models, we develop, for the first time, an analytical model that can capture the essential physical phenomena in a simplified yet representative manner for a biomimetic beam with uniformly spaced, rigid rectangular scales. Our objective is to highlight both the differences and similarities between this fluid-mediated viscoelastic dissipation and other fundamental damping mechanisms previously explored in these systems—such as Coulomb friction \cite{ali2019frictional}, air damping \cite{ali2019frictional}, and material viscosity \cite{ebrahimi2023material}. This model also lays the groundwork for investigating more complex configurations, including scale-covered plates and shells, functionally graded systems, and soft scales of varying shapes \cite{sarkar2025bending, ali2019tailorable}.

\section{System Configuration-- Geometry and Kinematics}

To begin our analysis, we first describe the inherently multiscale and multiphysics characteristics of this system, Fig.~\ref{6-Fig1}(b). At the largest length (O(m)) scale lies the overall structure, which comprises repeating unit cells (O(mm)), each formed by a scale in contact with its neighboring scale, with an interfacial region containing trapped fluid (O($\mu$m)). We model the scales as rigid and the trapped complex fluid as undergoing Couette flow between two flat plates confined within a narrow lubrication gap, Fig.~\ref{6-Fig1}(b). The speed of the Couette flow is taken to be the relative sliding velocity between two scales in contact, determined by the kinematics of the scales sliding. The total fluid-contact area is denoted by $a_c$, and the thickness of the hydrodynamic lubrication layer is $h$, Figs.~\ref{6-Fig1}(b). These quantities serve as system parameters and are assumed to remain constant during the entirety of the bending process. The actual lubrication can be a complex and transient combinations of Couette and squeeze film damping \cite{chong2019modelling}. However we assume here that the film thickness remains approximately constant, well formed with film disruption only occurring near the extremes of the bending loads where contact forces can rise sharply (locking). In practice this means lubricant (slime) is incompressible and deforms isothermally. We also assume that an appropriate amount of lubricant is available and able to flow into the contact region to maintain a constant, thin film thickness, and with negligible surface roughness effects \cite{bhushan2003introduction}. In addition, this also implies that the pressure gradient is small compared to the shear stress gradient. The scales are uniformly spaced with distance \( d \), have an exposed length \( l \), with scale width \(b\), and are embedded into the substrate to a depth \( L \). Each scale is inclined at an angle \( \theta \), with \( \theta_0 \) representing the initial inclination before engagement, Fig.~\ref{6-Fig1}(c). The scale thickness is denoted by \( D \), the substrate's Young's modulus by \( E_B \), and the area moment of the beam cross-section as $I_B$. We use a kinematic relationship among the scale angle \( \theta \), the substrate curvature angle \( \psi \), and the scale overlap ratio \( \eta = \frac{l}{d} \) from the literature of biomimetic beams\cite{ghosh2014contact}: $\theta = \sin^{-1} \left( \eta \psi \cos \frac{\psi}{2} \right) - \frac{\psi}{2}$. This relationship exhibits a singularity near a specific $\psi = \psi_{\text{lock}}$ explained earlier \cite{ghosh2014contact}. This locking phenomenon has emerged as a unique feature of the biomimetic scale system(s) that appears even under combination loads \cite{dharmavaram2022coupled}, non-periodic engagement \cite{ali2019bending, ali2019tailorable}, plate systems \cite{sarkar2025bending}, and even under non-ideal conditions \cite{ghosh2017non}. Taking the time derivative of this equation connects the angular velocity of scales $(\dot\theta)$ to the substrate curvature rate $(\dot\psi)$ yields the following angular velocity relationship:

{
\begin{equation}
\dot{\theta} = \dot{\psi} \frac{\partial\theta}{\partial\psi} = \dot{\psi} \left( \frac{\eta \left( \cos \frac{\psi}{2} - \frac{\psi}{2} \sin \frac{\psi}{2} \right)}{\sqrt{1 - \eta^2 \psi^2 \cos^2 \frac{\psi}{2}} - \frac{1}{2}} \right)
\end{equation}
}

This relationship shows a nonlinear scaling between the two angular velocities parameterized by the scale overlap ratio $\eta$.

To obtain the relative sliding velocity between the scales, we now focus on the next length scale, which is the unit cell consisting of a pair of sliding scales, Fig.~\ref{6-Fig1}(c). From the geometry of sliding at any given angle $\theta\ge\theta_0$, using the coordinates of point \( E \) = $\left(d - l \cos(\theta - \psi), \frac{\psi d}{2} + l \sin(\theta - \psi)\right)$ \cite{sarkar2025bending}, we get, $r=l \sqrt{\left(\frac{1}{\eta} - \cos(\theta - \psi)\right)^2 + \left(\frac{\psi}{2\eta} + \sin(\theta - \psi)\right)^2}$, where $\eta=l/d$ is the overlap ratio. Taking derivative with time, we get the sliding velocity as $v_{rel}=\dot{r}=\frac{\partial r}{\partial \psi} \dot \psi$.

\section{Mechanics: Elasticity of Bending and Hydrodynamics of Lubrication}

We first start our analysis at the smallest length scale, the interface between adjacent scales, Fig.~\ref{6-Fig1}(c), where the trapped fluid generates a resistive force given by \( F_D = \mu \frac{a_c}{h} v_{rel} \). Here, \( \mu \) denotes the fluid viscosity, and \( v_{rel} \) represents the relative sliding velocity between the scales. To model a complex fluid, we adopt the Carreau fluid model \cite{bowers2023modeling}, expressed as: \(\mu = \mu_\infty + (\mu_0 - \mu_\infty) \left( 1 + \frac{\Lambda^2}{h^2} \dot{r}^2 \right)^{\frac{m-1}{2}}\). In this formulation, \( \mu_\infty \) and \( \mu_0 \) correspond to the asymptotic and initial viscosities, respectively, while \( \Lambda \) is the characteristic time (s), and $\dot{r}$ is the relative sliding velocity between scales. The parameter \( m \) governs the degree of non-Newtonian behavior, with \( m < 1 \) representing shear thinning liquid, \( m > 1 \), shear thickening, and \( m = 1 \) recovering the Newtonian fluid case with $\mu = \mu_0$. Now, substituting the expression of $v_{rel}$ into the flow equation, we get $F_D=\mu\frac{a_c}{h}\dot\psi\frac{\partial r}{\partial \psi}$. From this relationship, it is clear that even for a Newtonian fluid, the linear relationship between force and velocity breaks down due to the nonlinear geometrical multiplier factor $\frac{\partial r}{\partial \psi}$. This can be non-dimensionalized as $\bar r'=\frac{1}{l}\frac{\partial r}{\partial \psi}$, and expressing the equation of dissipative force, $F_D=\mu\frac{a_c l}{h}\dot\psi\frac{\partial \bar{r}}{\partial \psi}$.





To analyze the mechanics of the system, we use an energy approach to couple the length scales. Here, the external work done due to the applied bending load on the beam per unit length is  $\frac{1}{d} \int_0^\psi M d\psi$. Elastic energy of the substrate deformation is $\frac{1}{2} E_B I_B \psi^2 \frac{1}{d^2}$, scale rotation energy per scale is $\frac{1}{2d} K_{\theta} (\theta - \theta _0)^2$, where $K_\theta$ is the base rotational stiffness opposing scales rotation. The dissipated energy due to fluid sliding $\frac{1}{d} \int_{r_e}^{r} F_{D} \, dr$. Thus, the balance of work-energies can be written as:

\begin{equation}
\begin{split}
\frac{1}{d} \int_{0}^{\psi} M \, d\psi = \frac{1}{2} E_B I_B \psi^2 \frac{1}{d^2} + \frac{1}{2d} K_{\theta} (\theta - \theta_0)^2 \cdot \mathcal{H}(\theta - \theta_0) \\
+ \frac{1}{d} \int_{r_e}^{r} F_{D} \, dr \cdot \mathcal{H}(\theta - \theta_0).
\end{split}
\label{6-Eq2}
\end{equation}

$\mathcal{H}(.)$ is the Heaviside step function that activates the nonlinear terms after engagement. For a more comprehensive analysis of the viscoelasticity, the energy dissipation due to the presence of Couette flow is calculated and compared with the total work done on the system: $W_{\text{sys}} = U_{\text{el}} + W_{\text{D}}$, where $U_{\text{el}}$ is the elastic energy of the system, and $W_{\text{D}}$ is the total dissipative work done from the Couette flow lubrication. The elastic energy of the system is:

\begin{equation}
U_{\text{el}} = \frac{1}{2} E_BI \psi^2 \frac{1}{d^2} + \frac{1}{2d} K_{\theta} (\theta - \theta_0)^2 \cdot \mathcal{H}(\theta - \theta_0).
\end{equation}

The total dissipative work done due to Couette flow is:

\begin{equation}
W_{\text{D}} = \frac{1}{d} \int_{r_e}^{r} F_{D} \, dr \cdot \mathcal{H}(\theta - \theta_0).
\end{equation}

We define the relative energy dissipation (RED) factor as the ratio of the dissipative work $W_{\text{D}}$, to the total work done on the system $W_{\text{sys}}$ as $\text{RED} = \frac{W_{\text{D}}}{W_{\text{sys}}}$. This fraction has been used to characterize dissipation of such systems earlier in literature \cite{ghosh2016frictional,ebrahimi2020coulomb}.

To derive the moment-curvature relationship, we differentiate Equation (\ref{6-Eq2}) with respect to \( \psi\). Normalizing the equation by dividing it by $\frac{E_BI_B}{d}$ and substituting the expressions of $I_B = \frac{1}{12} bH^3$ ($b$ is beam thickness, $H$ is the beam height), spring constant $K_\theta$ = $C_B E_B b D^2 \left( \frac{L}{D} \right)^n$ ($C_B$ = 0.66, and $n = 1.75$) \cite{ghosh2014contact}, and dissipative force $F_D=\mu\frac{a_c l}{h}\dot\psi\frac{\partial \bar{r}}{\partial \psi}$ in above equation:

\begin{equation}
\begin{split}
\Bar{M} = \psi + 12 C_B \left( \frac{L}{D} \right)^n \left( \frac{D}{H} \right)^2 \left( \frac{d}{H} \right) (\theta - \theta_0) \frac{\partial \theta}{\partial \psi} \cdot \mathcal{H}(\theta - \theta_0)\\
+ 12 \left[ \frac{\mu}{E_B} \dot{\psi} \right] \left( \frac{\alpha_L}{\delta_L} \right) \left( \frac{l}{H} \right)^3 \left(\frac{\partial\bar{r}}{\partial\psi}\right)^2 \cdot \mathcal{H}(\theta - \theta_0).
\end{split}
\end{equation}

where $\alpha_L = \frac{a_c}{bl}$ is the ratio of lubrication area to scale area, and $\delta_L = \frac{h}{d}$ is the ratio of lubrication gap to scale spacing. Note that this implies that $\alpha_L\delta_L = \dfrac{a_c h}{b l d}$, which is the volume fraction of the lubrication zone with respect to the total RVE volume. We neglect the inertia of the scales, assuming the mass of the scales is negligible compared to the mass of the substrate \cite{ali2019frictional}.

In this study, both $\delta_L$ and \( \alpha_L\) are taken as constant parameters during the deformation process. This assumption of $\delta_L$ is motivated by the typical characteristics of thin film lubrication, where the film thickness $h$ remains approximately steady over short time scales and small deformations, particularly when the elastic deformation of the substrate is negligible compared to the rigid scale motions \cite{tavakol2017extended, hamrock2004fundamentals}. In the present case, since the scales are modeled as rigid and smooth, and the relative motion occurs primarily through sliding rather than compression or separation, the variation in $h$ is minimal, justifying the approximation of constant $\delta_L$ \cite{elkholy2007granular}. The assumptions stated above also ensure that changes in the lubrication area $\alpha_L$ (contact patch) remain limited. From classical lubrication theory, specifically under the Couette flow regime, the shear stress and resulting resistive force are highly sensitive to the gap height $h$, which appears in the denominator of the viscous force term. If $h$ were to vary significantly, the viscous resistance $F_D \propto \mu a_c v_{\mathrm{rel}}/h$ would become nonlinear not only due to geometry changes or fluid rheology, but also because of variation in thickness, complicating the viscoelastic behavior \cite{tavakol2017extended, baric2016extended}. In practical terms, our simplifying assumptions correspond to an adequately coated or flooded lubrication condition, where the slime or fluid fills the interfacial contact zone uniformly across all scales \cite{chong2019modelling}. As a result, any variation in \( a_c \) due to local curvature-induced divergence is assumed negligible compared to the dominant contribution from overall scale geometry. Holding \( \delta_L \) and \( \alpha_L \) constant also allows us to isolate the influence of other critical parameters—such as fluid rheology and sliding kinematics—on energy dissipation, while still capturing the dominant effects of scale area and fluid-film confinement \cite{chong2019modelling}. Furthermore, when $h$ becomes too large (i.e., $\delta_L$ increases), the flow regime transitions from a confined lubrication regime to either partial or broken film conditions, where the fluid may no longer bridge the surfaces, resulting in loss of viscous coupling. This film rupture phenomenon occurs when surface separation exceeds a critical threshold beyond which capillary and van der Waals forces can no longer sustain the fluid bridge \cite{coyne1970conditions, ota1987note}. Considering such phenomena would require full hydrodynamic or elastohydrodynamic modeling \cite{hamrock2004fundamentals, nosov2004appearance}, which is outside the scope of this study.

Two different classes of imposed dynamic loads are considered for analysis: a ramp load with a constant ramp rate \( \dot{\psi} \), and an oscillatory load with frequencies ranging from subharmonic to superharmonic relative to the natural frequency of the underlying unscaled beam. In the oscillatory case, the curvature is prescribed as a sinusoidal function, \( \psi(t) = \psi_{\text{lock}} \sin(\Omega t) \). The applied frequency \( \Omega \) is a ratio of actual applied frequency to the natural frequency $\Omega_n$, defined as $\Omega_n = \pi^2 \sqrt{\frac{E_B I_B}{\rho_B A L_B^4}}$ \cite{alzghoul2022dynamic, blevins2015formulas}, where \( \rho_B \) is the density of the substrate material. Here, \( \psi_{\text{lock}} \) denotes the locking curvature of the beam, which is typically \( \psi_{\text{lock}} \) $\approx$ \( 1/\eta \) \cite{ghosh2014contact} for moderate to large $\eta$. However, in this study, we limit the analysis to \( \psi_{\text{lock}} \) $=$ \( 0.9/\eta \) to avoid singularity effects on elasticity and tribological assumptions.

\begin{figure*}[htbp]
\centering
\begin{tabular}{cc}
\includegraphics[scale = 0.85]{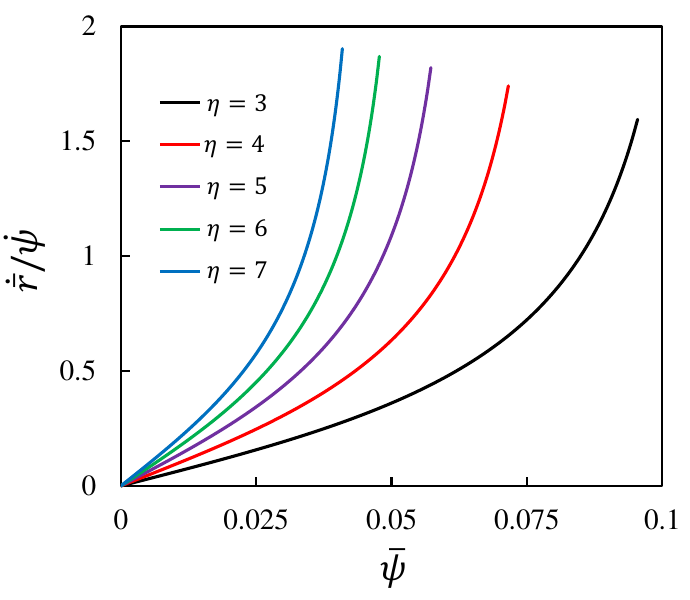} 
\end{tabular}
\caption{Variation of the normalized relative sliding velocity \( \dot{\bar{r}}/\dot{\psi} \) with \( \bar{\psi} \) ($\psi/\pi$) for different values of \( \eta \). Here, results are plotted for constant, $\dot\psi$ = 1 rad/s.
}
 \label{6-Fig2}
\end{figure*}

\begin{figure*}[htbp]
\centering
\begin{tabular}{cc}
\includegraphics[scale = 0.75]{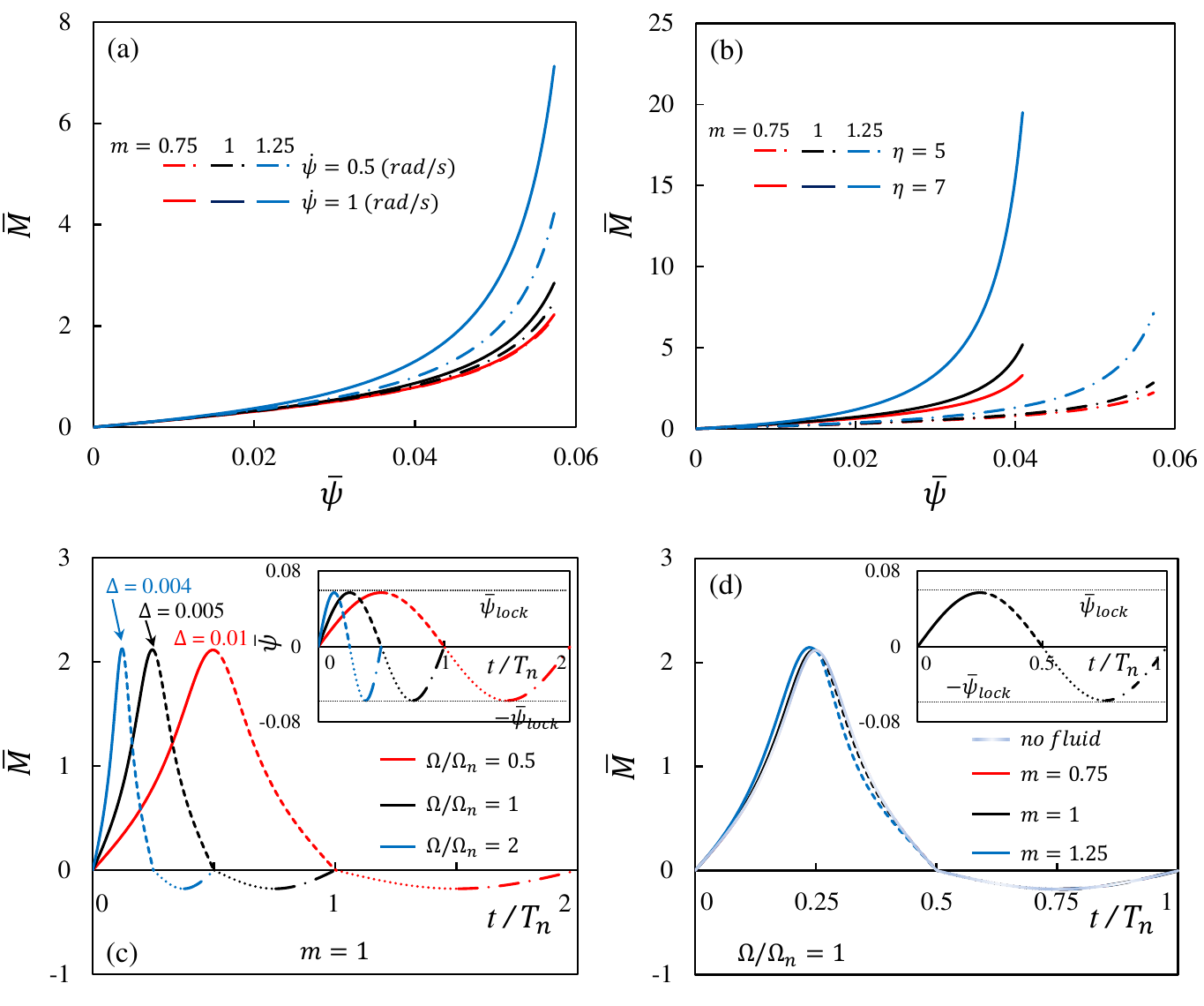} 
\end{tabular}
\caption{
(a) Variation of the normalized moment, \(\bar{M}\), with \(\bar{\psi}\) ($\psi/\pi$), for different values of \(m\) and \(\dot{\psi}\) ($\eta$ = 5).  
(b) Variation of the normalized moment, \(\bar{M}\), with \(\bar{\psi}\), for different values of \(m\) and \(\eta\) (\(\dot{\psi} = 1 \, \text{rad/s}\)). 
(c) Time-dependent variation of \(\bar{M}\) for different frequency ratios, \(\Omega/\Omega_n\). The inset highlights the variation of the bending curvature over time for different values of \(\Omega/\Omega_n\) ($\eta$ = 5).
(d) Time-dependent variation of the normalized moment, \(\bar{M}\), for different \(m\) values at \(\Omega/\Omega_n = 1\) ($\eta$ = 5). The inset highlights the variation of the bending curvature over time for \(\Omega/\Omega_n\) = 1.
Here, results are plotted for \(\delta_L = 2 \times 10^{-4}\) and \(\alpha_L = 0.01\), and $T_n$ in Fig. 3(c)-(d) is the time period corresponding to the natural frequency of the unscaled beam. 
}
 \label{6-Fig3}
\end{figure*}

\begin{figure*}[htbp]
\centering
\begin{tabular}{cc}
\includegraphics[scale = 0.75]{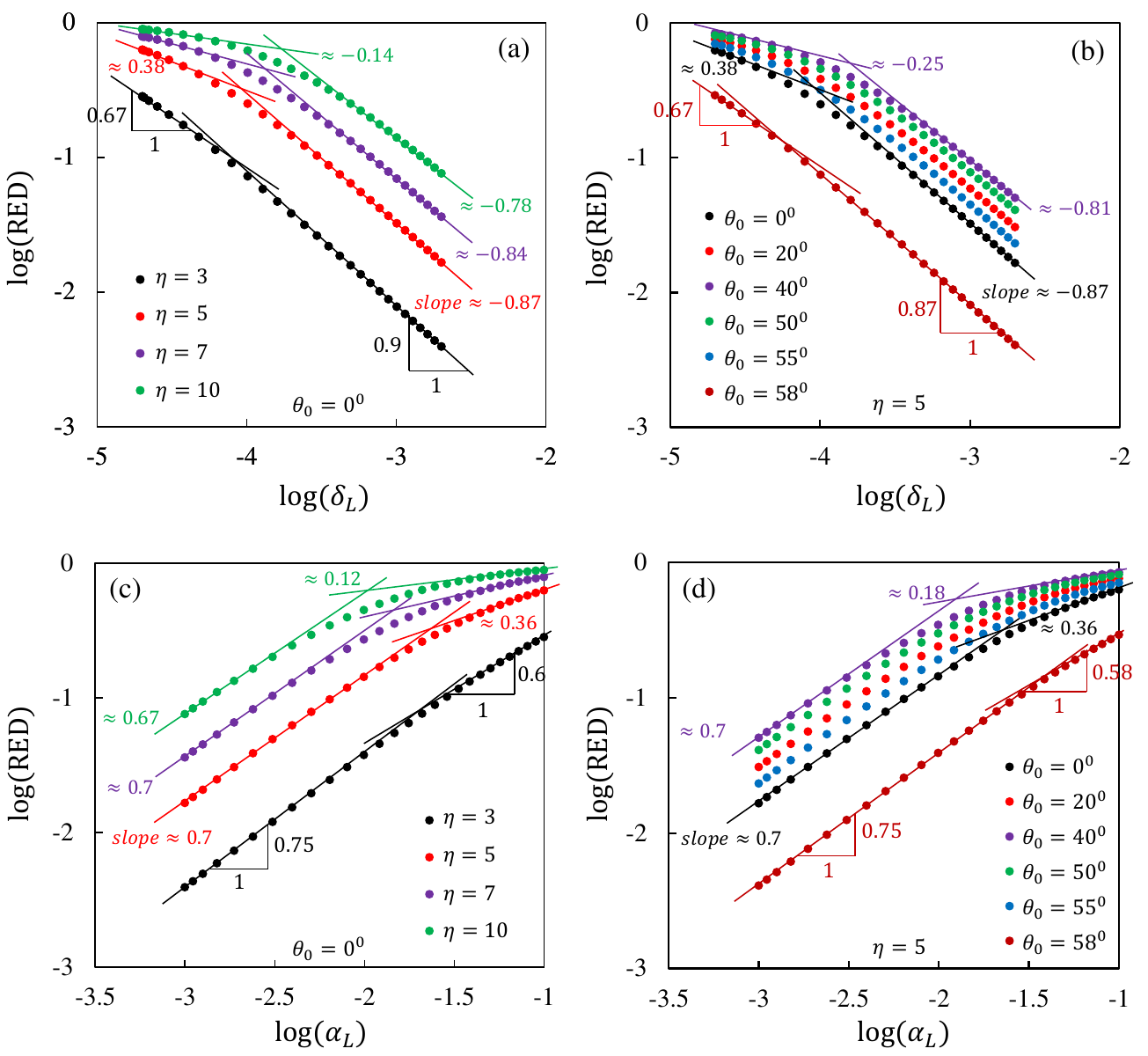} 
\end{tabular}
\caption{
Log--log plots of the Relative Energy Dissipation (RED) factor with respect to tribological parameters: (a-b) RED versus the lubrication gap \( \delta_L \) (with \( \alpha_L = 0.01 \)), for two cases: (a) varying overlap ratio \( \eta \), and (b) varying initial inclination angle \( \theta_0 \); (c-d) RED versus the lubrication area ratio \( \alpha_L \) (with \( \delta_L = 2 \times 10^{-4} \)), for two cases: (c) varying \( \eta \), and (d) varying \( \theta_0 \). The results correspond to \( \bar{E}_B = 3.4 \times 10^{-4} \), \( m = 1 \), and a constant curvature rate \( \dot{\psi} = 1 \, \text{rad/s} \).
}
 \label{6-Fig4}
\end{figure*}

\begin{figure*}[htbp]
\centering
\begin{tabular}{cc}
\includegraphics[scale = 0.75]{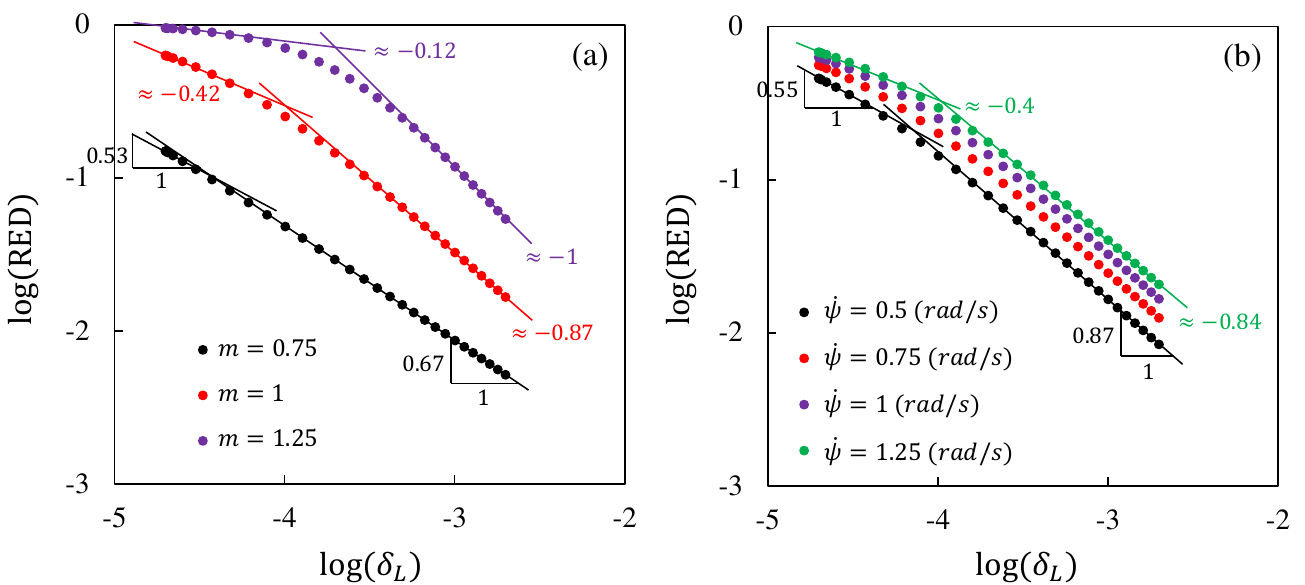} 
\end{tabular}
\caption{
Log--log plots of the RED factor with the lubrication gap parameter \( \delta_L \) varying: (a) \( m \) (\( \dot{\psi} = 1\, \mathrm{rad/s} \)), and (b) \( \dot{\psi} \) (\( m = 1 \)). The results correspond to \( \bar{E}_B = 3.4 \times 10^{-4} \), \( \eta = 5 \), \( \theta_0 = 0^\circ \), and \( \alpha_L = 0.01 \).
}
 \label{6-Fig5}
\end{figure*}

\begin{figure*}[htbp]
\centering
\begin{tabular}{cc}
\includegraphics[scale = 1.1]{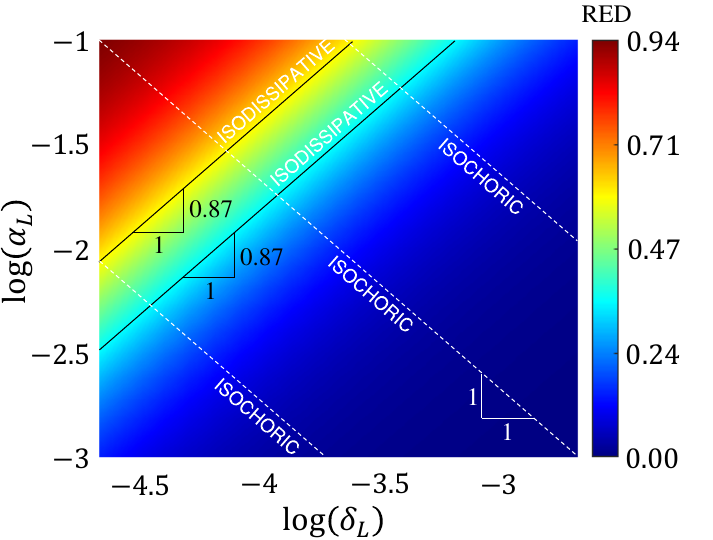} 
\end{tabular}
\caption{Contour plot of the RED factor as a function of the lubrication gap parameter $\delta_L$ and the lubrication area ratio $\alpha_L$, illustrating their combined influence on energy dissipation. The solid black lines represent empirically observed isodissipation contours, indicating that RED remains approximately constant along curves where $\delta_L \cdot \alpha_L^{0.87} = \text{constant}$. The white dashed lines denote isochoric lines corresponding to $log(\delta_L) + log(\alpha_L) = \text{constant}$, corresponding to the constant volume fractions of the complex fluid. Results are plotted for $\bar{E}_B = 3.4 \times 10^{-4}$, $\eta = 5$, $\theta_0 = 0^\circ$, $m$ = 1, and a constant curvature rate $\dot{\psi} = 1~\mathrm{rad/s}$. }
\label{6-Fig6}
\end{figure*}

\begin{figure*}[htbp]
\centering
\begin{tabular}{cc}
\includegraphics[scale = 0.76]{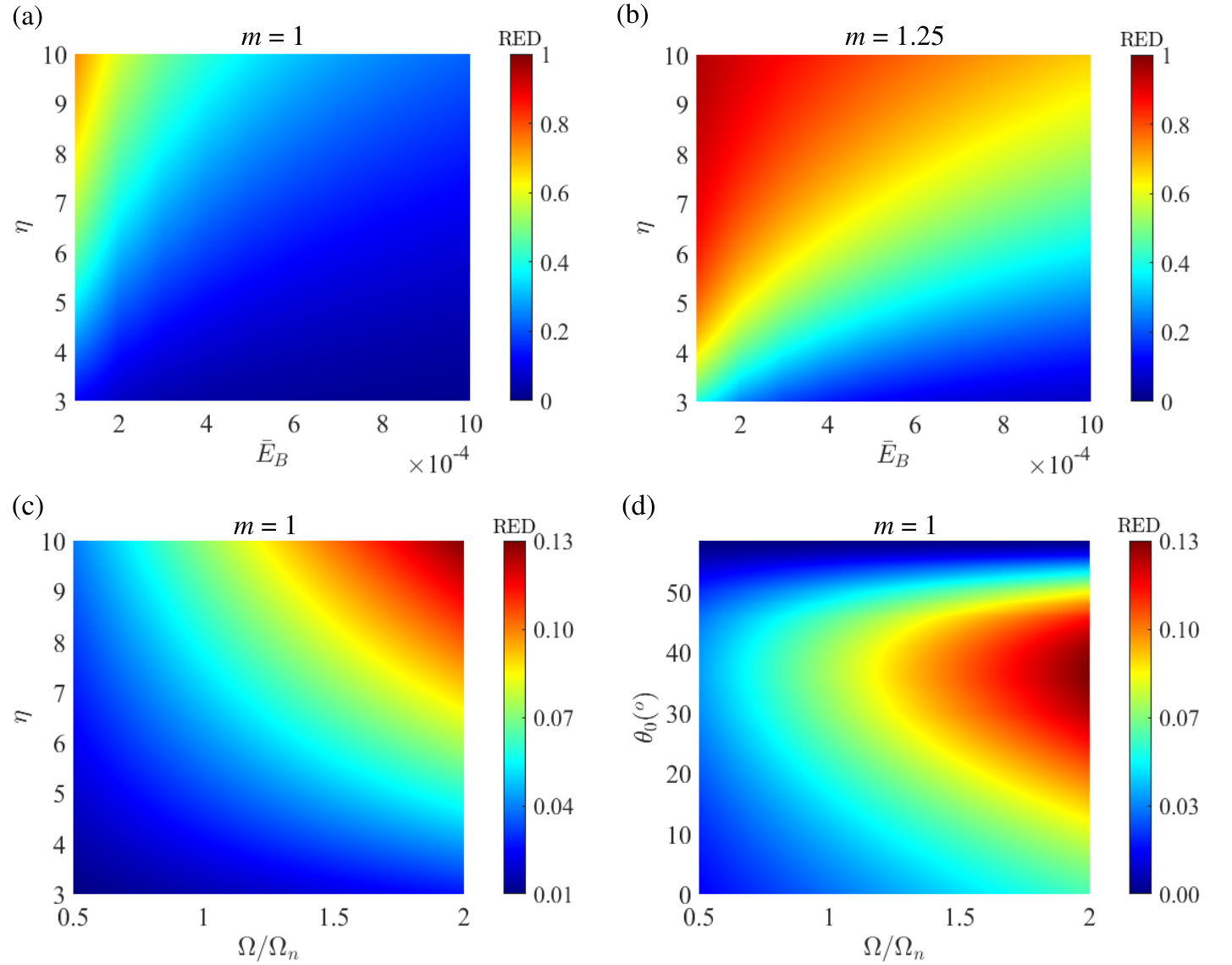} 
\end{tabular}
\caption{Non-dimensional Relative Energy Dissipation (RED) factor phase plots spanned by: (a-b) \( \eta \) and \( \bar{E}_B \) for two different values of \( m \): (a) \( m = 1 \), (b) \( m = 1.25 \). In Figures (a)-(b) results are plotted for a constant bending curvature rate $\dot\psi = 1$ rad/s, $\alpha_L = 0.01$, $\delta_L = 2 \times 10^{-4}$, and $\theta_0 = 0^o$. 
(c) Contour plot of RED spanned by the overlap ratio $\eta$ and \(\Omega/\Omega_n\) (\( \bar{E}_B = 3.4 \times 10^{-4} \), $\alpha_L = 0.01$, $\delta_L = 2 \times 10^{-4}$, $\theta_0 = 0^o$). (d) Contour plot of RED spanned by $\theta_0$ and \(\Omega/\Omega_n\)($\eta = 5$, \( \bar{E}_B = 3.4 \times 10^{-4} \), $\alpha_L = 0.01$, $\delta_L = 2 \times 10^{-4}$). In Figures (c)-(d), results are plotted up to the first quarter of the cyclic load.
}
 \label{6-Fig7}
\end{figure*}

\section{Results and Discussion}

For simulations, we consider a rectangular substrate with a length \( L_B = 64 \, \text{mm} \). Note that we do not require explicit values for the beam or scale width, as the analysis is performed using non-dimensional quantities. Throughout the analysis, the longitudinal distance between scales, \( d \), is kept constant at 5 mm, with an embedded scale length \( L = 1 \, \text{mm} \) and scale thickness \( D = 0.1 \, \text{mm} \). Initially, the exposed length of the scale is set to \( l = 25 \, \text{mm} \), resulting in \( \eta = 5 \), with the initial scale inclination angle \( \theta_0 = 0^\circ \). \( \alpha_L \) is fixed at 0.01, and the lubrication gap is set to \( h = 1 \, \mu\text{m} \), resulting in \( \delta_L = 2 \times 10^{-4}\). Although $\delta_L$ and $\alpha_L$ are held constant unless stated otherwise,  we still carry out simulations showing the effect of varying $\delta_L$ and $\alpha_L$. The initial and  asymptotic viscosity values are considered as \({\mu_0} = 1\) Pa.s, \({\mu_\infty} = 0.01\) Pa.s, respectively, and \( \Lambda = 0.05 \) s. In this study, the scale spacing is kept constant at \( d = 5 \, \text{mm} \), while the exposed scale length \( l \) and lubrication gap \( h \) are varied to vary the parameter values of \( \eta \) and \( \delta_{L} \), respectively. The elastic modulus of the substrate is taken as \( E_B = 0.34 \, \text{MPa} \), while that of the scale is \( E_{sc} = 10^3 \, \text{MPa} \). These stiffnesses correspond to the high contrasts between soft substrates (polymeric) and stiffer mineralized scales in nature. In artificial systems, these differences capture soft rubbery (silicone) materials and stiffer thermoplastics or metals that can be used to make scales. This high contrast in elastic modulus between the substrate and the scales also justifies the assumption of rigid scales. 


We first illustrate the intrinsic kinematic nonlinearity of the system in Fig.~\ref{6-Fig2}, which presents the normalized sliding velocity ratio, $\dot{\bar{r}}/\dot{\psi}$, as a function of normalized curvature, $\bar{\psi}$ ($\psi/\pi$), for various overlap ratios $\eta$. The plot reveals exponential dependence of sliding velocity ratios on imposed curvature for any value of the overlap ratio $\eta$. This kinematic amplification of sliding ensures that as nonlinear elasticity rises due to scale engagement, so does the fluidic resistance. This implies that internal fluid force, which is proportional to the local fluid velocity gradient, can increase sharply due to purely geometric global-local amplification even when the underlying fluid is Newtonian, thereby introducing a new form of viscoelasticity in such systems. Furthermore, this amplification leads to an increase in the viscous loss component alongside the nonlinear elastic response.

This nonlinear amplification of fluid dissipative force arising from the nonlinear sliding velocity is reflected in the moment–curvature plots in Fig.~\ref{6-Fig3}(a), which show a pronounced strain-rate sensitivity across all three classes of fluids considered (shear-thinning, Newtonian, and shear-thickening), when all other geometrical and tribological parameters are held constant. The figure confirms that fluid rheology plays a significant role in influencing the strain-rate sensitivity of the bending response, with shear-thinning liquids showing much lower sensitivity than shear-thickening ones. Fig.~\ref{6-Fig3}(a) also allows us to draw a comparative analysis with the standard viscoelastic models (standard linear models) of solids. Such a comparison is best done using a qualitative analysis of this system. For our purpose, we use a relaxation test where a sudden strain is applied to the system and then held constant, while we observe the load behavior. This relaxation test can be envisioned as a sudden high ramp rate load, followed by a quick succession of strain rate decreases that stabilize the strain to a constant value. Under this relaxation load protocol, the moment evolution can be envisioned from Fig.~\ref{6-Fig3}(a). At the end of loading at a given strain rate, the moment is at its maximum. Then, the strain rate goes to zero rapidly, which can be envisioned as a series of rapid successive falls of moment magnitude through progressively lower strain rates (initial moment point falling through the lower strain rate curves in the Fig.~\ref{6-Fig3}(a)), until at zero strain rate, the moment is same as the elastic case (no more fluid dissipation). This relaxation behavior is qualitatively similar to the Zener model of linear viscoelasticity (system with one-spring element is attached in parallel to a Maxwell element) \cite{christensen2013theory}, although our system is nonlinear. Note that in addition to strain rates, even $\eta$ will have a significant impact on the moment-curvature relationship, as evidenced by the kinematic velocity relationships shown in Fig.~\ref{6-Fig2}. To isolate this synergistic behavior, Fig.~\ref{6-Fig3}(b) presents moment–curvature plots at a constant strain rate (1/s) for two different overlap ratios. The plot shows that the strain-rate sensitivity induced by fluid rheology is considerably amplified at higher overlaps. Hence, the viscoelastic response arises from both the fluid’s rheology and the scale overlap, both acting synergistically with each other. We now investigate the effect of sinusoidal loads on the dynamic behavior of the beam, keeping all other material and geometric parameters constant. The normalized time is defined as \( t/T_n \), where \( T_n = \frac{2\pi}{\Omega_n} \) corresponds to the natural period of the underlying unscaled beam. The dynamic load is prescribed as curvature \( \bar\psi \) $(\psi/\pi)$ that varies from 0 to the locking curvature $\bar\psi_{lock}$ \( (= 0.9/\eta\pi) \) in both directions during one cycle. We first study the moment evolution with time under the imposed sinusoidal load at different frequencies. These responses for a Newtonian fluid are shown in Fig.~\ref{6-Fig3}(c). Two aspects are of interest in this plot. First, the asymmetry of the substrate (scales only on one side) translates into asymmetry in the moment response. Second, the nonlinearity in one direction distorts the shape of the moment curve over time, making typical linear viscoelasticity frameworks based on modulis less meaningful. A more comprehensive approach would therefore be to study the overall energy profile, particularly the relationship between elastic and viscous energy contributions from the fluid. Despite this, the moment curves still exhibit characteristics reminiscent of linear viscoelasticity, for instance, the diminishing time lag between the peaks of moment and curvature with increasing frequency. This points toward greater solid-like (glassy) behavior of the underlying substrate at higher frequencies, similar to that observed in linear viscoelastic systems. Next, we fix the frequency and observe the effect of fluid rheology in Fig.~\ref{6-Fig3}(d). We find that the moment peaks in the shear-thickening regime are noticeably higher and show the most phase lag relative to the curvature. In contrast, the response for shear-thinning fluids closely resembles the elastic case. Although the moment waveforms are distorted due to nonlinearity and asymmetry (scales only on one side), the behavior is consistent with viscoelasticity, where stress typically leads strain under dynamic loading. These interplays between elastic and viscous components are better understood in terms of relative energy ratios (RED factor introduced earlier). These serve as a generalization of the ratios of loss and storage moduli ($G''$, $G'$) in linear viscoelasticity. The behavior of these moduli (energies in our case) are considered to reflect the underlying microstructural mechanisms of the deformation of these materials.


The RED factor captures the dissipative or loss component of viscoelasticity and, in general, depends on underlying mechanisms determined by elastic properties, fluid rheology, and tribological parameters. Note that these represent all three length scales delineated in Fig.~\ref{6-Fig1}(b). RED factor also captures the interplay between the elastic forces of the beam and viscous forces of the lubricant. If the ratio of elastic to viscous work is denoted by $\beta$, we can see RED = 1/(1+$\beta$). Hence, $\beta$ and RED are inversely correlated. This ratio appears analogously in complex fluid literature (Weissenberg number, $Wi$, and Deborah number, $De$, for viscoelastic fluids), and liquid crystals (Ericksen number, $Er$). They all come to signify similar competition but with differing contexts. In liquid crystals, the fluid actively deforms the underlying structure (director field), and higher Er signifies more pronounced fluidic effects. However, in this case, the fluid characteristics have a negligible effect on the solid. Wi and De are more relevant in this context, with the difference that the elasticity in the current system comes from the scales, while in complex viscoelastic fluids assumed elasticity is due to elongated molecules or polymers dispersed in the fluid itself. We begin our investigation by isolating parameters at the smallest length scale—the tribological scale. Here, the two most important factors are the lubrication gap and the lubrication area. The dependence on both of these parameters is best studied using a log-log scale. For brevity, we focus only on Newtonian fluids for these plots. Fig.~\ref{6-Fig4}(a) plots the RED dependence on the lubrication gap \( \delta_L \) for various overlap ratios. It reveals two important characteristics of the dissipation behavior. First, the RED factor exhibits a power-law dependence on the lubrication gap, and second, a regime-differentiated dissipation behavior emerges, separating the low and high gap regimes. At lower gaps, the power-law exponent is greater for lower \( \eta \) and smaller for higher \( \eta \). Thus, at higher \( \eta \), dissipation is less sensitive to the lubrication gap. Interestingly, this plot also shows that as the gap increases, the power-law curves converge to a similar exponent (between 0.8 and 1) over a range of \( \eta \). These exponents suggest that at higher \( \delta_L \), dissipation drops off rapidly. Hence, two distinct lubrication regimes emerge: a less sensitive regime at lower gaps and a more sensitive, rapidly decaying regime at higher gaps. The dependence of dissipation on another geometric parameter, \( \theta_0 \) (initial inclination), is even more interesting. Similar to the lubrication gap, at higher \( \delta_L \), there is again a constant power-law dependence between RED and the gap. But, unlike the trend with \( \eta \), the effect of initial angle shows a markedly different progression. Regime differentiation becomes stronger as \( \theta_0 \) increases, similar to increasing \( \eta \). However, this trend holds only up to a certain initial angle. Beyond this value, increasing \( \theta_0 \) begins to flatten the RED curve, reducing the starkness of the regime differentiation and eventually almost eliminating it at high enough angles. Next, we look at the effect of the lubrication area $(\alpha_L)$ on the dissipation, Figs.~\ref{6-Fig4}(c) - ~\ref{6-Fig4}(d). Here too, power-law and regime-differentiated dissipation behavior is evident. Just like lubrication gap, here also higher overlap produces a pronounced regime differentiated behavior with increasing $\alpha_L$ (Fig.~\ref{6-Fig4}(c)). Lower lubrication area also removes the sensitivity of the overall power law exponent on $\eta$, showing that in the low $\alpha_L$, the dependence on $\eta$ comes undone. Fig.~\ref{6-Fig4}(d) shows the dissipation variation with $\alpha_L$, with different initial angles. Here also, the behavior is similar to Fig.~\ref{6-Fig4}(b), where, beyond topical similarity to $\eta$ variation, there is an intermediate value of initial angle where regime differentiation is most pronounced.

In Fig.~\ref{6-Fig5}(a), we assess whether the previously observed dual-regime trend is sensitive to the rheology of the fluid. The results indicate that shear-thickening (\( m = 1.25 \)) significantly amplifies dissipation in the strong lubrication regime, leading to a steeper RED slope at lower \( \delta_L \). Conversely, shear-thinning fluids (\( m = 0.75 \)) display a more gradual decline in RED, confirming that fluid rheology plays a critical role in modulating dissipation, especially under a high degree of lubricant confinement. The effect of imposed strain rate is also studied next in Fig.~\ref{6-Fig5}(b). Here, we find that higher strain rates decrease the sensitivity of dissipation to the lubrication gap at lower values of \( \delta_L \), while showing minimal influence in the higher gap regime. Note that, as expected, \( \alpha_L \) exhibits a similar but opposite (increasing) trend to that of \( \delta_L \) with varying \( m \) and \( \dot{\psi} \), consistent with the behaviors shown in Figs.~\ref{6-Fig4}(c) - ~\ref{6-Fig4}(d). Therefore, the corresponding plots are omitted here for brevity.

To summarize the overall dependence of RED on the tribological parameters, we develop a phase plot where RED is spanned by both \( \delta_L \) and \( \alpha_L \) on a logarithmic scale for a representative Newtonian fluid, as shown in Fig.~\ref{6-Fig6}. This phase plot comes with an intrinsic level set that is defined by the straight lines \( \log(\alpha_L) + \log(\delta_L) = \text{constant} \), which are essentially isochoric lines with constant lubricant volume. These are shown as white dashed lines in Fig.~\ref{6-Fig6}. In addition, once the phase plot is developed, another family of level sets—the lines connecting the same dissipation level- can also be found. Interestingly, these also show straight-line contours with a constant slope (0.87). This indicates that for any level of dissipation, \( \delta_L \cdot \alpha_L^{0.87} = \text{constant} \). It is conceivable that this scaling exponent would vary with rheological and geometrical parameters. These isodissipation lines (black solid) are also shown in Fig.~\ref{6-Fig6}. This plot highlights the nonlinear nature of RED variation—very high dissipation at lower \( \delta_L \). Thus, for a constant volume of lubricant, the dissipation increases much more effectively when \( \delta_L \) is lowered than when \( \alpha_L \) is increased.

We now carry out further analysis of the dissipation on various system and load parameters in greater detail, fixing $\alpha_L$ = 0.01, and $\delta_L = 2\times10^{-4}$, which is representative of the strong lubrication regime. Figs.~\ref{6-Fig7}(a)-~\ref{6-Fig7}(b) presents RED factor spanned by the overlap ratio \(\eta\) and the modulus ratio \(\bar{E}_B\) ($E_{B}$/$E_{sc}$), for Newtonian (\( m = 1 \)) and shear-thickening (\( m = 1.25 \)) fluids, respectively. Here, \( E_B \) is varied while keeping the scale modulus fixed at \( E_{sc} = 10^3 \)~MPa. In both cases, increasing substrate stiffness \( E_B \) leads to greater elastic energy storage. At the same time, increasing \( \eta \) enhances both stiffness (due to scale rotation) and damping (due to increased sliding velocities, Fig.~\ref{6-Fig2}). These competing effects lead to interesting dynamics: while elasticity grows with \( E_B \), the damping contribution from \( \eta \) becomes more dominant at higher overlap values. This interplay is evident in Fig.~\ref{6-Fig7}(a), where the slopes of iso-damping lines (\( \Delta \eta / \Delta \bar{E}_B \)) decrease as both \( \bar{E}_B \) and \( \eta \) increase, indicating that much larger increases in stiffness are required to maintain the same dissipation level. These effects are even more pronounced for shear-thickening fluids, as shown in Fig.~\ref{6-Fig7}(b). Therefore, any fluid trapped between the scales, particularly Newtonian or shear-thickening, can make high-density scale systems inherently dissipative. In such cases, higher density scales typically contribute more to damping than to elasticity, effectively dominating the energy dissipation behavior of the structure. Extending our investigation into sinusoidal loads, we now consider dissipation variation as a phase plot mapped by scale overlap and applied frequency. This is depicted in Fig.~\ref{6-Fig7}(c), which shows that a relatively lower overlap ratio, frequency has little effect. However, at the higher overlaps, increasing frequencies can lead to substantially higher dissipation. Finally, we consider the effect of the initial scale inclination angle $\theta_0$. This is another crucial variable that has been shown to have a substantial effect on the nonlinear elasticity of the substrates \cite{ali2019tailorable}. In Figs.~\ref{6-Fig4}(b) and 4(d), we saw that RED shows a pronounced effect with the increase of \( \theta_0 \), but after a certain value of \( \theta_0 \), the RED curves start to fall-off, which alludes to a non-monotonic dependence of RED on $\theta_0$. To investigate this effect in greater depth, in Fig.~\ref{6-Fig7}(d), we plotted the RED contour plot spanned by \( \Omega/\Omega_n \) and \( \theta_0 \). Here, the effect of increasing $\theta_0$ would be to delay the onset of engagement and hence damping. However, when engagement does occur, dissipation starts at a higher value due to the high velocity of sliding at engagement that operates at a higher curvature, Fig.~\ref{6-Fig2}. At the same time, late engagement also leads to lowered elastic contribution from scales rotation. Thus, we would expect that dissipation would increase with higher angles. This is in general true, as seen in Fig.~\ref{6-Fig7}(d). However,  paradoxically, relative dissipation starts to drop off after intermediate values of initial inclination angle. This is true even for higher frequencies, where dissipation is, in general, higher. This points to two competing factors at play. The first factor is the increase in starting dissipation at higher angles, but when angles are substantially high, the total range of curvatures through which dissipation can act also reduces. Thus, in the maximum case, there $\theta_0$ = $90^0$, the total dissipation would be zero since no engagement took place at all before locking. This phenomena is analogous to the effect of Coulomb friction on RED, which is known to exhibit maximum dissipation at intermediate values of friction coefficients (angle of friction). As before, this is a result of competing factors that arise as angle of friction is raised - increased friction force leading to higher moment, while reduction in range of motion due to advancing the locking curvature\cite{ghosh2016frictional}. Interestingly, in this case, the rheology does not qualitatively change the overall phase plot with respect to $\theta_0$, merely changing the magnitude of the observed dissipation and hence not shown here for brevity.



\section{Conclusion}

We develop, for the first time, the theoretical foundations to investigate the nonlinear viscoelastic response of a biomimetic scale-covered beam embedded with shear-dependent complex fluids trapped between scales. The study highlights the interplay of system geometry, tribological parameters, and fluid rheology that shapes the emergent viscoelasticity. By studying both constant strain rate and oscillatory loading, we identify a combination of universal and diverse phenomena intrinsic to the dissipative behavior of this class of systems. Our analysis reveals that overlapping scales work synergistically with fluid rheology, leading to emergent nonlinearity in viscoelastic behavior. These nonlinear effects are studied using the concept of relative energy dissipation (RED), defined as the ratio of dissipation to total energy, and acts as an analog to the loss modulus for linear viscoelasticity. RED factor analysis reveals the regime-differentiated nature of the dissipation with respect to tribological parameters, a differentiation that is further modulated by both geometric and rheological properties of the trapped complex fluid. We also report that greater scale overlap and shear-thickening behavior enhance the viscous component of the response, while increased substrate stiffness tends to suppresses it. In this context, even though increased overlap is the primary factor for nonlinear strain stiffening of the substrate, increased shear thickening amplifies its role as a dissipator, overpowering the elastic contribution. Moreover, the RED factor’s dependence on initial scale inclination exposes a nontraditional and dual nature that leads to maximizing dissipation at intermediate scale angles. These insights point toward a new class of tunable, adaptive structures where fluid-mediated viscoelasticity can be engineered for targeted performance - offering design pathways for soft robotics, aerospace morphing surfaces, and protective materials requiring on-demand modulation of stiffness and dissipation.


\section*{Acknowledgement}
This work was supported by the United States National
Science Foundation’s Civil, Mechanical, and Manufacturing
Innovation, Grant No. 2028338. We
gratefully acknowledge the support of European Research Council through Starting Grant MUCUS (grant
no. ERC-StG-2019-852529).


\section*{Data Availability Statement}

\appendix

\section{Appendixes}

\nocite{*}
\bibliography{aipsamp}

\begin{thebibliography}{60}%
\makeatletter
\providecommand \@ifxundefined [1]{%
 \@ifx{#1\undefined}
}%
\providecommand \@ifnum [1]{%
 \ifnum #1\expandafter \@firstoftwo
 \else \expandafter \@secondoftwo
 \fi
}%
\providecommand \@ifx [1]{%
 \ifx #1\expandafter \@firstoftwo
 \else \expandafter \@secondoftwo
 \fi
}%
\providecommand \natexlab [1]{#1}%
\providecommand \enquote  [1]{``#1''}%
\providecommand \bibnamefont  [1]{#1}%
\providecommand \bibfnamefont [1]{#1}%
\providecommand \citenamefont [1]{#1}%
\providecommand \href@noop [0]{\@secondoftwo}%
\providecommand \href [0]{\begingroup \@sanitize@url \@href}%
\providecommand \@href[1]{\@@startlink{#1}\@@href}%
\providecommand \@@href[1]{\endgroup#1\@@endlink}%
\providecommand \@sanitize@url [0]{\catcode `\\12\catcode `\$12\catcode `\&12\catcode `\#12\catcode `\^12\catcode `\_12\catcode `\%12\relax}%
\providecommand \@@startlink[1]{}%
\providecommand \@@endlink[0]{}%
\providecommand \url  [0]{\begingroup\@sanitize@url \@url }%
\providecommand \@url [1]{\endgroup\@href {#1}{\urlprefix }}%
\providecommand \urlprefix  [0]{URL }%
\providecommand \Eprint [0]{\href }%
\providecommand \doibase [0]{http://dx.doi.org/}%
\providecommand \selectlanguage [0]{\@gobble}%
\providecommand \bibinfo  [0]{\@secondoftwo}%
\providecommand \bibfield  [0]{\@secondoftwo}%
\providecommand \translation [1]{[#1]}%
\providecommand \BibitemOpen [0]{}%
\providecommand \bibitemStop [0]{}%
\providecommand \bibitemNoStop [0]{.\EOS\space}%
\providecommand \EOS [0]{\spacefactor3000\relax}%
\providecommand \BibitemShut  [1]{\csname bibitem#1\endcsname}%
\let\auto@bib@innerbib\@empty
\bibitem [{\citenamefont {Lovegrove}(2001)}]{lovegrove2001evolution}%
  \BibitemOpen
  \bibfield  {author} {\bibinfo {author} {\bibfnamefont {B.~G.}\ \bibnamefont {Lovegrove}},\ }\bibfield  {title} {\enquote {\bibinfo {title} {The evolution of body armor in mammals: plantigrade constraints of large body size},}\ }\href@noop {} {\bibfield  {journal} {\bibinfo  {journal} {Evolution}\ }\textbf {\bibinfo {volume} {55}},\ \bibinfo {pages} {1464--1473} (\bibinfo {year} {2001})}\BibitemShut {NoStop}%
\bibitem [{\citenamefont {Bruet}\ \emph {et~al.}(2008)\citenamefont {Bruet}, \citenamefont {Song}, \citenamefont {Boyce},\ and\ \citenamefont {Ortiz}}]{bruet2008materials}%
  \BibitemOpen
  \bibfield  {author} {\bibinfo {author} {\bibfnamefont {B.~J.}\ \bibnamefont {Bruet}}, \bibinfo {author} {\bibfnamefont {J.}~\bibnamefont {Song}}, \bibinfo {author} {\bibfnamefont {M.~C.}\ \bibnamefont {Boyce}}, \ and\ \bibinfo {author} {\bibfnamefont {C.}~\bibnamefont {Ortiz}},\ }\bibfield  {title} {\enquote {\bibinfo {title} {Materials design principles of ancient fish armour},}\ }\href@noop {} {\bibfield  {journal} {\bibinfo  {journal} {Nature materials}\ }\textbf {\bibinfo {volume} {7}},\ \bibinfo {pages} {748--756} (\bibinfo {year} {2008})}\BibitemShut {NoStop}%
\bibitem [{\citenamefont {Yang}\ \emph {et~al.}(2013)\citenamefont {Yang}, \citenamefont {Chen}, \citenamefont {Gludovatz}, \citenamefont {Zimmermann}, \citenamefont {Ritchie},\ and\ \citenamefont {Meyers}}]{yang2013natural}%
  \BibitemOpen
  \bibfield  {author} {\bibinfo {author} {\bibfnamefont {W.}~\bibnamefont {Yang}}, \bibinfo {author} {\bibfnamefont {I.~H.}\ \bibnamefont {Chen}}, \bibinfo {author} {\bibfnamefont {B.}~\bibnamefont {Gludovatz}}, \bibinfo {author} {\bibfnamefont {E.~A.}\ \bibnamefont {Zimmermann}}, \bibinfo {author} {\bibfnamefont {R.~O.}\ \bibnamefont {Ritchie}}, \ and\ \bibinfo {author} {\bibfnamefont {M.~A.}\ \bibnamefont {Meyers}},\ }\bibfield  {title} {\enquote {\bibinfo {title} {Natural flexible dermal armor},}\ }\href@noop {} {\bibfield  {journal} {\bibinfo  {journal} {Advanced Materials}\ }\textbf {\bibinfo {volume} {25}},\ \bibinfo {pages} {31--48} (\bibinfo {year} {2013})}\BibitemShut {NoStop}%
\bibitem [{\citenamefont {Gower}(2003)}]{gower2003scale}%
  \BibitemOpen
  \bibfield  {author} {\bibinfo {author} {\bibfnamefont {D.~J.}\ \bibnamefont {Gower}},\ }\bibfield  {title} {\enquote {\bibinfo {title} {Scale microornamentation of uropeltid snakes},}\ }\href@noop {} {\bibfield  {journal} {\bibinfo  {journal} {Journal of Morphology}\ }\textbf {\bibinfo {volume} {258}},\ \bibinfo {pages} {249--268} (\bibinfo {year} {2003})}\BibitemShut {NoStop}%
\bibitem [{\citenamefont {Prum}, \citenamefont {Quinn},\ and\ \citenamefont {Torres}(2006)}]{prum2006anatomically}%
  \BibitemOpen
  \bibfield  {author} {\bibinfo {author} {\bibfnamefont {R.~O.}\ \bibnamefont {Prum}}, \bibinfo {author} {\bibfnamefont {T.}~\bibnamefont {Quinn}}, \ and\ \bibinfo {author} {\bibfnamefont {R.~H.}\ \bibnamefont {Torres}},\ }\bibfield  {title} {\enquote {\bibinfo {title} {Anatomically diverse butterfly scales all produce structural colours by coherent scattering},}\ }\href@noop {} {\bibfield  {journal} {\bibinfo  {journal} {Journal of Experimental Biology}\ }\textbf {\bibinfo {volume} {209}},\ \bibinfo {pages} {748--765} (\bibinfo {year} {2006})}\BibitemShut {NoStop}%
\bibitem [{\citenamefont {Doucet}\ and\ \citenamefont {Meadows}(2009)}]{doucet2009iridescence}%
  \BibitemOpen
  \bibfield  {author} {\bibinfo {author} {\bibfnamefont {S.~M.}\ \bibnamefont {Doucet}}\ and\ \bibinfo {author} {\bibfnamefont {M.~G.}\ \bibnamefont {Meadows}},\ }\bibfield  {title} {\enquote {\bibinfo {title} {Iridescence: a functional perspective},}\ }\href@noop {} {\bibfield  {journal} {\bibinfo  {journal} {Journal of the Royal Society Interface}\ }\textbf {\bibinfo {volume} {6}},\ \bibinfo {pages} {S115--S132} (\bibinfo {year} {2009})}\BibitemShut {NoStop}%
\bibitem [{\citenamefont {Denton}(1970)}]{denton1970review}%
  \BibitemOpen
  \bibfield  {author} {\bibinfo {author} {\bibfnamefont {E.~J.}\ \bibnamefont {Denton}},\ }\bibfield  {title} {\enquote {\bibinfo {title} {Review lecture: on the organization of reflecting surfaces in some marine animals},}\ }\href@noop {} {\bibfield  {journal} {\bibinfo  {journal} {Philosophical Transactions of the Royal Society of London. B, Biological Sciences}\ }\textbf {\bibinfo {volume} {258}},\ \bibinfo {pages} {285--313} (\bibinfo {year} {1970})}\BibitemShut {NoStop}%
\bibitem [{\citenamefont {Drucker}\ and\ \citenamefont {Lauder}(2002)}]{drucker2002experimental}%
  \BibitemOpen
  \bibfield  {author} {\bibinfo {author} {\bibfnamefont {E.~G.}\ \bibnamefont {Drucker}}\ and\ \bibinfo {author} {\bibfnamefont {G.~V.}\ \bibnamefont {Lauder}},\ }\bibfield  {title} {\enquote {\bibinfo {title} {Experimental hydrodynamics of fish locomotion: functional insights from wake visualization},}\ }\href@noop {} {\bibfield  {journal} {\bibinfo  {journal} {Integrative and Comparative Biology}\ }\textbf {\bibinfo {volume} {42}},\ \bibinfo {pages} {243--257} (\bibinfo {year} {2002})}\BibitemShut {NoStop}%
\bibitem [{\citenamefont {Borazjani}\ and\ \citenamefont {Sotiropoulos}(2008)}]{borazjani2008numerical}%
  \BibitemOpen
  \bibfield  {author} {\bibinfo {author} {\bibfnamefont {I.}~\bibnamefont {Borazjani}}\ and\ \bibinfo {author} {\bibfnamefont {F.}~\bibnamefont {Sotiropoulos}},\ }\bibfield  {title} {\enquote {\bibinfo {title} {Numerical investigation of the hydrodynamics of carangiform swimming in the transitional and inertial flow regimes},}\ }\href@noop {} {\bibfield  {journal} {\bibinfo  {journal} {Journal of experimental biology}\ }\textbf {\bibinfo {volume} {211}},\ \bibinfo {pages} {1541--1558} (\bibinfo {year} {2008})}\BibitemShut {NoStop}%
\bibitem [{\citenamefont {Oeffner}\ and\ \citenamefont {Lauder}(2012)}]{oeffner2012hydrodynamic}%
  \BibitemOpen
  \bibfield  {author} {\bibinfo {author} {\bibfnamefont {J.}~\bibnamefont {Oeffner}}\ and\ \bibinfo {author} {\bibfnamefont {G.~V.}\ \bibnamefont {Lauder}},\ }\bibfield  {title} {\enquote {\bibinfo {title} {The hydrodynamic function of shark skin and two biomimetic applications},}\ }\href@noop {} {\bibfield  {journal} {\bibinfo  {journal} {Journal of Experimental Biology}\ }\textbf {\bibinfo {volume} {215}},\ \bibinfo {pages} {785--795} (\bibinfo {year} {2012})}\BibitemShut {NoStop}%
\bibitem [{\citenamefont {Chintapalli}\ \emph {et~al.}(2014)\citenamefont {Chintapalli}, \citenamefont {Mirkhalaf}, \citenamefont {Dastjerdi},\ and\ \citenamefont {Barthelat}}]{chintapalli2014fabrication}%
  \BibitemOpen
  \bibfield  {author} {\bibinfo {author} {\bibfnamefont {R.~K.}\ \bibnamefont {Chintapalli}}, \bibinfo {author} {\bibfnamefont {M.}~\bibnamefont {Mirkhalaf}}, \bibinfo {author} {\bibfnamefont {A.~K.}\ \bibnamefont {Dastjerdi}}, \ and\ \bibinfo {author} {\bibfnamefont {F.}~\bibnamefont {Barthelat}},\ }\bibfield  {title} {\enquote {\bibinfo {title} {Fabrication, testing and modeling of a new flexible armor inspired from natural fish scales and osteoderms},}\ }\href@noop {} {\bibfield  {journal} {\bibinfo  {journal} {Bioinspiration \& biomimetics}\ }\textbf {\bibinfo {volume} {9}},\ \bibinfo {pages} {036005} (\bibinfo {year} {2014})}\BibitemShut {NoStop}%
\bibitem [{\citenamefont {Martini}, \citenamefont {Balit},\ and\ \citenamefont {Barthelat}(2017)}]{martini2017comparative}%
  \BibitemOpen
  \bibfield  {author} {\bibinfo {author} {\bibfnamefont {R.}~\bibnamefont {Martini}}, \bibinfo {author} {\bibfnamefont {Y.}~\bibnamefont {Balit}}, \ and\ \bibinfo {author} {\bibfnamefont {F.}~\bibnamefont {Barthelat}},\ }\bibfield  {title} {\enquote {\bibinfo {title} {A comparative study of bio-inspired protective scales using 3d printing and mechanical testing},}\ }\href@noop {} {\bibfield  {journal} {\bibinfo  {journal} {Acta biomaterialia}\ }\textbf {\bibinfo {volume} {55}},\ \bibinfo {pages} {360--372} (\bibinfo {year} {2017})}\BibitemShut {NoStop}%
\bibitem [{\citenamefont {Martini}\ and\ \citenamefont {Barthelat}(2016)}]{martini2016stretch}%
  \BibitemOpen
  \bibfield  {author} {\bibinfo {author} {\bibfnamefont {R.}~\bibnamefont {Martini}}\ and\ \bibinfo {author} {\bibfnamefont {F.}~\bibnamefont {Barthelat}},\ }\bibfield  {title} {\enquote {\bibinfo {title} {Stretch-and-release fabrication, testing and optimization of a flexible ceramic armor inspired from fish scales},}\ }\href@noop {} {\bibfield  {journal} {\bibinfo  {journal} {Bioinspiration \& biomimetics}\ }\textbf {\bibinfo {volume} {11}},\ \bibinfo {pages} {066001} (\bibinfo {year} {2016})}\BibitemShut {NoStop}%
\bibitem [{\citenamefont {Ibanez}, \citenamefont {Cowx},\ and\ \citenamefont {O'HIGGINS}(2009)}]{ibanez2009variation}%
  \BibitemOpen
  \bibfield  {author} {\bibinfo {author} {\bibfnamefont {A.~L.}\ \bibnamefont {Ibanez}}, \bibinfo {author} {\bibfnamefont {I.~G.}\ \bibnamefont {Cowx}}, \ and\ \bibinfo {author} {\bibfnamefont {P.}~\bibnamefont {O'HIGGINS}},\ }\bibfield  {title} {\enquote {\bibinfo {title} {Variation in elasmoid fish scale patterns is informative with regard to taxon and swimming mode},}\ }\href@noop {} {\bibfield  {journal} {\bibinfo  {journal} {Zoological Journal of the Linnean Society}\ }\textbf {\bibinfo {volume} {155}},\ \bibinfo {pages} {834--844} (\bibinfo {year} {2009})}\BibitemShut {NoStop}%
\bibitem [{\citenamefont {Rudykh}, \citenamefont {Ortiz},\ and\ \citenamefont {Boyce}(2015)}]{rudykh2015flexibility}%
  \BibitemOpen
  \bibfield  {author} {\bibinfo {author} {\bibfnamefont {S.}~\bibnamefont {Rudykh}}, \bibinfo {author} {\bibfnamefont {C.}~\bibnamefont {Ortiz}}, \ and\ \bibinfo {author} {\bibfnamefont {M.~C.}\ \bibnamefont {Boyce}},\ }\bibfield  {title} {\enquote {\bibinfo {title} {Flexibility and protection by design: imbricated hybrid microstructures of bio-inspired armor},}\ }\href@noop {} {\bibfield  {journal} {\bibinfo  {journal} {Soft Matter}\ }\textbf {\bibinfo {volume} {11}},\ \bibinfo {pages} {2547--2554} (\bibinfo {year} {2015})}\BibitemShut {NoStop}%
\bibitem [{\citenamefont {Fudge}\ \emph {et~al.}(2005)\citenamefont {Fudge}, \citenamefont {Levy}, \citenamefont {Chiu},\ and\ \citenamefont {Gosline}}]{fudge2005composition}%
  \BibitemOpen
  \bibfield  {author} {\bibinfo {author} {\bibfnamefont {D.~S.}\ \bibnamefont {Fudge}}, \bibinfo {author} {\bibfnamefont {N.}~\bibnamefont {Levy}}, \bibinfo {author} {\bibfnamefont {S.}~\bibnamefont {Chiu}}, \ and\ \bibinfo {author} {\bibfnamefont {J.~M.}\ \bibnamefont {Gosline}},\ }\bibfield  {title} {\enquote {\bibinfo {title} {Composition, morphology and mechanics of hagfish slime},}\ }\href@noop {} {\bibfield  {journal} {\bibinfo  {journal} {Journal of Experimental Biology}\ }\textbf {\bibinfo {volume} {208}},\ \bibinfo {pages} {4613--4625} (\bibinfo {year} {2005})}\BibitemShut {NoStop}%
\bibitem [{\citenamefont {Wainwright}, \citenamefont {Lauder},\ and\ \citenamefont {Gemmell}(2024)}]{wainwright2024hydrodynamic}%
  \BibitemOpen
  \bibfield  {author} {\bibinfo {author} {\bibfnamefont {D.~K.}\ \bibnamefont {Wainwright}}, \bibinfo {author} {\bibfnamefont {G.~V.}\ \bibnamefont {Lauder}}, \ and\ \bibinfo {author} {\bibfnamefont {B.~J.}\ \bibnamefont {Gemmell}},\ }\bibfield  {title} {\enquote {\bibinfo {title} {Hydrodynamic function of the slimy and scaly surfaces of teleost fishes},}\ }\href@noop {} {\bibfield  {journal} {\bibinfo  {journal} {Integrative and Comparative Biology}\ }\textbf {\bibinfo {volume} {64}},\ \bibinfo {pages} {480--495} (\bibinfo {year} {2024})}\BibitemShut {NoStop}%
\bibitem [{\citenamefont {Fischer}, \citenamefont {Lauder},\ and\ \citenamefont {Wainwright}(2025)}]{fischer2025slippery}%
  \BibitemOpen
  \bibfield  {author} {\bibinfo {author} {\bibfnamefont {M.~J.}\ \bibnamefont {Fischer}}, \bibinfo {author} {\bibfnamefont {G.~V.}\ \bibnamefont {Lauder}}, \ and\ \bibinfo {author} {\bibfnamefont {D.~K.}\ \bibnamefont {Wainwright}},\ }\bibfield  {title} {\enquote {\bibinfo {title} {Slippery and smooth shark skin: How mucus transforms surface texture},}\ }\href@noop {} {\bibfield  {journal} {\bibinfo  {journal} {Journal of Morphology}\ }\textbf {\bibinfo {volume} {286}},\ \bibinfo {pages} {e70046} (\bibinfo {year} {2025})}\BibitemShut {NoStop}%
\bibitem [{\citenamefont {Wainwright}\ and\ \citenamefont {Lauder}(2017)}]{wainwright2017mucus}%
  \BibitemOpen
  \bibfield  {author} {\bibinfo {author} {\bibfnamefont {D.~K.}\ \bibnamefont {Wainwright}}\ and\ \bibinfo {author} {\bibfnamefont {G.~V.}\ \bibnamefont {Lauder}},\ }\bibfield  {title} {\enquote {\bibinfo {title} {Mucus matters: the slippery and complex surfaces of fish},}\ }\href@noop {} {\bibfield  {journal} {\bibinfo  {journal} {Functional Surfaces in Biology III: Diversity of the Physical Phenomena}\ ,\ \bibinfo {pages} {223--246}} (\bibinfo {year} {2017})}\BibitemShut {NoStop}%
\bibitem [{\citenamefont {Chaudhary}, \citenamefont {Ewoldt},\ and\ \citenamefont {Thiffeault}(2019)}]{chaudhary2019unravelling}%
  \BibitemOpen
  \bibfield  {author} {\bibinfo {author} {\bibfnamefont {G.}~\bibnamefont {Chaudhary}}, \bibinfo {author} {\bibfnamefont {R.~H.}\ \bibnamefont {Ewoldt}}, \ and\ \bibinfo {author} {\bibfnamefont {J.-L.}\ \bibnamefont {Thiffeault}},\ }\bibfield  {title} {\enquote {\bibinfo {title} {Unravelling hagfish slime},}\ }\href@noop {} {\bibfield  {journal} {\bibinfo  {journal} {Journal of the Royal Society Interface}\ }\textbf {\bibinfo {volume} {16}},\ \bibinfo {pages} {20180710} (\bibinfo {year} {2019})}\BibitemShut {NoStop}%
\bibitem [{\citenamefont {Rashad}\ \emph {et~al.}(2023)\citenamefont {Rashad}, \citenamefont {Samp{\`o}}, \citenamefont {Cataldi},\ and\ \citenamefont {Zara}}]{rashad2023biological}%
  \BibitemOpen
  \bibfield  {author} {\bibinfo {author} {\bibfnamefont {M.}~\bibnamefont {Rashad}}, \bibinfo {author} {\bibfnamefont {S.}~\bibnamefont {Samp{\`o}}}, \bibinfo {author} {\bibfnamefont {A.}~\bibnamefont {Cataldi}}, \ and\ \bibinfo {author} {\bibfnamefont {S.}~\bibnamefont {Zara}},\ }\bibfield  {title} {\enquote {\bibinfo {title} {Biological activities of gastropods secretions: Snail and slug slime},}\ }\href@noop {} {\bibfield  {journal} {\bibinfo  {journal} {Natural Products and Bioprospecting}\ }\textbf {\bibinfo {volume} {13}},\ \bibinfo {pages} {42} (\bibinfo {year} {2023})}\BibitemShut {NoStop}%
\bibitem [{\citenamefont {Barajas-Ledesma}\ and\ \citenamefont {Holland}(2023)}]{barajas2023probing}%
  \BibitemOpen
  \bibfield  {author} {\bibinfo {author} {\bibfnamefont {E.}~\bibnamefont {Barajas-Ledesma}}\ and\ \bibinfo {author} {\bibfnamefont {C.}~\bibnamefont {Holland}},\ }\bibfield  {title} {\enquote {\bibinfo {title} {Probing the compositional and rheological properties of gastropod locomotive mucus},}\ }\href@noop {} {\bibfield  {journal} {\bibinfo  {journal} {Frontiers in Soft Matter}\ }\textbf {\bibinfo {volume} {3}},\ \bibinfo {pages} {1201511} (\bibinfo {year} {2023})}\BibitemShut {NoStop}%
\bibitem [{\citenamefont {Yan}\ \emph {et~al.}(2022)\citenamefont {Yan}, \citenamefont {Gu}, \citenamefont {Ma}, \citenamefont {Tang}, \citenamefont {He}, \citenamefont {Zhang},\ and\ \citenamefont {Mou}}]{yan2022slime}%
  \BibitemOpen
  \bibfield  {author} {\bibinfo {author} {\bibfnamefont {M.}~\bibnamefont {Yan}}, \bibinfo {author} {\bibfnamefont {Y.}~\bibnamefont {Gu}}, \bibinfo {author} {\bibfnamefont {L.}~\bibnamefont {Ma}}, \bibinfo {author} {\bibfnamefont {J.}~\bibnamefont {Tang}}, \bibinfo {author} {\bibfnamefont {C.}~\bibnamefont {He}}, \bibinfo {author} {\bibfnamefont {J.}~\bibnamefont {Zhang}}, \ and\ \bibinfo {author} {\bibfnamefont {J.}~\bibnamefont {Mou}},\ }\bibfield  {title} {\enquote {\bibinfo {title} {Slime-groove drag reduction characteristics and mechanism of marine biomimetic surface},}\ }\href@noop {} {\bibfield  {journal} {\bibinfo  {journal} {Applied bionics and biomechanics}\ }\textbf {\bibinfo {volume} {2022}},\ \bibinfo {pages} {4485365} (\bibinfo {year} {2022})}\BibitemShut {NoStop}%
\bibitem [{\citenamefont {Siddiqui}, \citenamefont {Burchard},\ and\ \citenamefont {Schwarz}(2001)}]{siddiqui2001undulating}%
  \BibitemOpen
  \bibfield  {author} {\bibinfo {author} {\bibfnamefont {A.}~\bibnamefont {Siddiqui}}, \bibinfo {author} {\bibfnamefont {R.}~\bibnamefont {Burchard}}, \ and\ \bibinfo {author} {\bibfnamefont {W.}~\bibnamefont {Schwarz}},\ }\bibfield  {title} {\enquote {\bibinfo {title} {An undulating surface model for the motility of bacteria gliding on a layer of non-newtonian slime},}\ }\href@noop {} {\bibfield  {journal} {\bibinfo  {journal} {International journal of non-linear mechanics}\ }\textbf {\bibinfo {volume} {36}},\ \bibinfo {pages} {743--761} (\bibinfo {year} {2001})}\BibitemShut {NoStop}%
\bibitem [{\citenamefont {Ali}\ \emph {et~al.}(2016)\citenamefont {Ali}, \citenamefont {Asghar}, \citenamefont {B{\'e}g},\ and\ \citenamefont {Sajid}}]{ali2016bacterial}%
  \BibitemOpen
  \bibfield  {author} {\bibinfo {author} {\bibfnamefont {N.}~\bibnamefont {Ali}}, \bibinfo {author} {\bibfnamefont {Z.}~\bibnamefont {Asghar}}, \bibinfo {author} {\bibfnamefont {O.~A.}\ \bibnamefont {B{\'e}g}}, \ and\ \bibinfo {author} {\bibfnamefont {M.}~\bibnamefont {Sajid}},\ }\bibfield  {title} {\enquote {\bibinfo {title} {Bacterial gliding fluid dynamics on a layer of non-newtonian slime: perturbation and numerical study},}\ }\href@noop {} {\bibfield  {journal} {\bibinfo  {journal} {Journal of theoretical biology}\ }\textbf {\bibinfo {volume} {397}},\ \bibinfo {pages} {22--32} (\bibinfo {year} {2016})}\BibitemShut {NoStop}%
\bibitem [{\citenamefont {Mahomed}\ \emph {et~al.}(2007)\citenamefont {Mahomed}, \citenamefont {Hayat}, \citenamefont {Momoniat},\ and\ \citenamefont {Asghar}}]{mahomed2007gliding}%
  \BibitemOpen
  \bibfield  {author} {\bibinfo {author} {\bibfnamefont {F.}~\bibnamefont {Mahomed}}, \bibinfo {author} {\bibfnamefont {T.}~\bibnamefont {Hayat}}, \bibinfo {author} {\bibfnamefont {E.}~\bibnamefont {Momoniat}}, \ and\ \bibinfo {author} {\bibfnamefont {S.}~\bibnamefont {Asghar}},\ }\bibfield  {title} {\enquote {\bibinfo {title} {Gliding motion of bacterium in a non-newtonian slime},}\ }\href@noop {} {\bibfield  {journal} {\bibinfo  {journal} {Nonlinear Analysis: Real World Applications}\ }\textbf {\bibinfo {volume} {8}},\ \bibinfo {pages} {853--864} (\bibinfo {year} {2007})}\BibitemShut {NoStop}%
\bibitem [{\citenamefont {Ewoldt}\ and\ \citenamefont {Saengow}(2022)}]{ewoldt2022designing}%
  \BibitemOpen
  \bibfield  {author} {\bibinfo {author} {\bibfnamefont {R.~H.}\ \bibnamefont {Ewoldt}}\ and\ \bibinfo {author} {\bibfnamefont {C.}~\bibnamefont {Saengow}},\ }\bibfield  {title} {\enquote {\bibinfo {title} {Designing complex fluids},}\ }\href@noop {} {\bibfield  {journal} {\bibinfo  {journal} {Annual Review of Fluid Mechanics}\ }\textbf {\bibinfo {volume} {54}},\ \bibinfo {pages} {413--441} (\bibinfo {year} {2022})}\BibitemShut {NoStop}%
\bibitem [{\citenamefont {Bowers}\ and\ \citenamefont {Miller}(2023)}]{bowers2023modeling}%
  \BibitemOpen
  \bibfield  {author} {\bibinfo {author} {\bibfnamefont {C.~A.}\ \bibnamefont {Bowers}}\ and\ \bibinfo {author} {\bibfnamefont {C.~T.}\ \bibnamefont {Miller}},\ }\bibfield  {title} {\enquote {\bibinfo {title} {Modeling flow of carreau fluids in porous media},}\ }\href@noop {} {\bibfield  {journal} {\bibinfo  {journal} {Physical Review E}\ }\textbf {\bibinfo {volume} {108}},\ \bibinfo {pages} {065106} (\bibinfo {year} {2023})}\BibitemShut {NoStop}%
\bibitem [{\citenamefont {Mandel}(2020)}]{mandel2020understanding}%
  \BibitemOpen
  \bibfield  {author} {\bibinfo {author} {\bibfnamefont {S.}~\bibnamefont {Mandel}},\ }\bibfield  {title} {\enquote {\bibinfo {title} {Understanding the power of mucus to reduce drag},}\ }\href@noop {} {\bibfield  {journal} {\bibinfo  {journal} {Scilight}\ }\textbf {\bibinfo {volume} {2020}} (\bibinfo {year} {2020})}\BibitemShut {NoStop}%
\bibitem [{\citenamefont {Wang}\ \emph {et~al.}(2020)\citenamefont {Wang}, \citenamefont {Ryu}, \citenamefont {He}, \citenamefont {Qin},\ and\ \citenamefont {Sung}}]{wang2020self}%
  \BibitemOpen
  \bibfield  {author} {\bibinfo {author} {\bibfnamefont {S.}~\bibnamefont {Wang}}, \bibinfo {author} {\bibfnamefont {J.}~\bibnamefont {Ryu}}, \bibinfo {author} {\bibfnamefont {G.-Q.}\ \bibnamefont {He}}, \bibinfo {author} {\bibfnamefont {F.}~\bibnamefont {Qin}}, \ and\ \bibinfo {author} {\bibfnamefont {H.~J.}\ \bibnamefont {Sung}},\ }\bibfield  {title} {\enquote {\bibinfo {title} {A self-propelled flexible plate with a navier slip surface},}\ }\href@noop {} {\bibfield  {journal} {\bibinfo  {journal} {Physics of Fluids}\ }\textbf {\bibinfo {volume} {32}} (\bibinfo {year} {2020})}\BibitemShut {NoStop}%
\bibitem [{\citenamefont {Sadati}\ \emph {et~al.}(2015)\citenamefont {Sadati}, \citenamefont {Noh}, \citenamefont {Elnaz~Naghibi}, \citenamefont {Althoefer},\ and\ \citenamefont {Nanayakkara}}]{sadati2015stiffness}%
  \BibitemOpen
  \bibfield  {author} {\bibinfo {author} {\bibfnamefont {S.~H.}\ \bibnamefont {Sadati}}, \bibinfo {author} {\bibfnamefont {Y.}~\bibnamefont {Noh}}, \bibinfo {author} {\bibfnamefont {S.}~\bibnamefont {Elnaz~Naghibi}}, \bibinfo {author} {\bibfnamefont {K.}~\bibnamefont {Althoefer}}, \ and\ \bibinfo {author} {\bibfnamefont {T.}~\bibnamefont {Nanayakkara}},\ }\bibfield  {title} {\enquote {\bibinfo {title} {Stiffness control of soft robotic manipulator for minimally invasive surgery (mis) using scale jamming},}\ }in\ \href@noop {} {\emph {\bibinfo {booktitle} {Intelligent Robotics and Applications: 9th International Conference, ICIRA 2015, Portsmouth, UK, August 24--27, 2015, Proceedings, Part III}}}\ (\bibinfo {organization} {Springer},\ \bibinfo {year} {2015})\ pp.\ \bibinfo {pages} {141--151}\BibitemShut {NoStop}%
\bibitem [{\citenamefont {Sire}, \citenamefont {Donoghue},\ and\ \citenamefont {Vickaryous}(2009)}]{sire2009origin}%
  \BibitemOpen
  \bibfield  {author} {\bibinfo {author} {\bibfnamefont {J.-Y.}\ \bibnamefont {Sire}}, \bibinfo {author} {\bibfnamefont {P.~C.}\ \bibnamefont {Donoghue}}, \ and\ \bibinfo {author} {\bibfnamefont {M.~K.}\ \bibnamefont {Vickaryous}},\ }\bibfield  {title} {\enquote {\bibinfo {title} {Origin and evolution of the integumentary skeleton in non-tetrapod vertebrates},}\ }\href@noop {} {\bibfield  {journal} {\bibinfo  {journal} {Journal of anatomy}\ }\textbf {\bibinfo {volume} {214}},\ \bibinfo {pages} {409--440} (\bibinfo {year} {2009})}\BibitemShut {NoStop}%
\bibitem [{\citenamefont {Ghosh}, \citenamefont {Ebrahimi},\ and\ \citenamefont {Vaziri}(2014)}]{ghosh2014contact}%
  \BibitemOpen
  \bibfield  {author} {\bibinfo {author} {\bibfnamefont {R.}~\bibnamefont {Ghosh}}, \bibinfo {author} {\bibfnamefont {H.}~\bibnamefont {Ebrahimi}}, \ and\ \bibinfo {author} {\bibfnamefont {A.}~\bibnamefont {Vaziri}},\ }\bibfield  {title} {\enquote {\bibinfo {title} {Contact kinematics of biomimetic scales},}\ }\href@noop {} {\bibfield  {journal} {\bibinfo  {journal} {Applied Physics Letters}\ }\textbf {\bibinfo {volume} {105}} (\bibinfo {year} {2014})}\BibitemShut {NoStop}%
\bibitem [{\citenamefont {Ebrahimi}\ \emph {et~al.}(2019)\citenamefont {Ebrahimi}, \citenamefont {Ali}, \citenamefont {Horton}, \citenamefont {Galvez}, \citenamefont {Gordon},\ and\ \citenamefont {Ghosh}}]{ebrahimi2019tailorable}%
  \BibitemOpen
  \bibfield  {author} {\bibinfo {author} {\bibfnamefont {H.}~\bibnamefont {Ebrahimi}}, \bibinfo {author} {\bibfnamefont {H.}~\bibnamefont {Ali}}, \bibinfo {author} {\bibfnamefont {R.~A.}\ \bibnamefont {Horton}}, \bibinfo {author} {\bibfnamefont {J.}~\bibnamefont {Galvez}}, \bibinfo {author} {\bibfnamefont {A.~P.}\ \bibnamefont {Gordon}}, \ and\ \bibinfo {author} {\bibfnamefont {R.}~\bibnamefont {Ghosh}},\ }\bibfield  {title} {\enquote {\bibinfo {title} {Tailorable twisting of biomimetic scale-covered substrate},}\ }\href@noop {} {\bibfield  {journal} {\bibinfo  {journal} {Europhysics Letters}\ }\textbf {\bibinfo {volume} {127}},\ \bibinfo {pages} {24002} (\bibinfo {year} {2019})}\BibitemShut {NoStop}%
\bibitem [{\citenamefont {Ghosh}, \citenamefont {Ebrahimi},\ and\ \citenamefont {Vaziri}(2016)}]{ghosh2016frictional}%
  \BibitemOpen
  \bibfield  {author} {\bibinfo {author} {\bibfnamefont {R.}~\bibnamefont {Ghosh}}, \bibinfo {author} {\bibfnamefont {H.}~\bibnamefont {Ebrahimi}}, \ and\ \bibinfo {author} {\bibfnamefont {A.}~\bibnamefont {Vaziri}},\ }\bibfield  {title} {\enquote {\bibinfo {title} {Frictional effects in biomimetic scales engagement},}\ }\href@noop {} {\bibfield  {journal} {\bibinfo  {journal} {EPL (Europhysics Letters)}\ }\textbf {\bibinfo {volume} {113}},\ \bibinfo {pages} {34003} (\bibinfo {year} {2016})}\BibitemShut {NoStop}%
\bibitem [{\citenamefont {Ebrahimi}, \citenamefont {Ali},\ and\ \citenamefont {Ghosh}(2020)}]{ebrahimi2020coulomb}%
  \BibitemOpen
  \bibfield  {author} {\bibinfo {author} {\bibfnamefont {H.}~\bibnamefont {Ebrahimi}}, \bibinfo {author} {\bibfnamefont {H.}~\bibnamefont {Ali}}, \ and\ \bibinfo {author} {\bibfnamefont {R.}~\bibnamefont {Ghosh}},\ }\bibfield  {title} {\enquote {\bibinfo {title} {Coulomb friction in twisting of biomimetic scale-covered substrate},}\ }\href@noop {} {\bibfield  {journal} {\bibinfo  {journal} {Bioinspiration \& biomimetics}\ }\textbf {\bibinfo {volume} {15}},\ \bibinfo {pages} {056013} (\bibinfo {year} {2020})}\BibitemShut {NoStop}%
\bibitem [{\citenamefont {Tatari}\ \emph {et~al.}(2023)\citenamefont {Tatari}, \citenamefont {Ebrahimi}, \citenamefont {Ghosh}, \citenamefont {Vaziri},\ and\ \citenamefont {Nayeb-Hashemi}}]{tatari2023bending}%
  \BibitemOpen
  \bibfield  {author} {\bibinfo {author} {\bibfnamefont {M.}~\bibnamefont {Tatari}}, \bibinfo {author} {\bibfnamefont {H.}~\bibnamefont {Ebrahimi}}, \bibinfo {author} {\bibfnamefont {R.}~\bibnamefont {Ghosh}}, \bibinfo {author} {\bibfnamefont {A.}~\bibnamefont {Vaziri}}, \ and\ \bibinfo {author} {\bibfnamefont {H.}~\bibnamefont {Nayeb-Hashemi}},\ }\bibfield  {title} {\enquote {\bibinfo {title} {Bending stiffness tunability of biomimetic scale covered surfaces via scales orientations},}\ }\href@noop {} {\bibfield  {journal} {\bibinfo  {journal} {International Journal of Solids and Structures}\ }\textbf {\bibinfo {volume} {280}},\ \bibinfo {pages} {112406} (\bibinfo {year} {2023})}\BibitemShut {NoStop}%
\bibitem [{\citenamefont {{Reefs.com Editorial Team}}(2020)}]{Reefs2020}%
  \BibitemOpen
  \bibfield  {author} {\bibinfo {author} {\bibnamefont {{Reefs.com Editorial Team}}},\ }\href@noop {} {\enquote {\bibinfo {title} {Slime: An aquarist’s best friend},}\ }\bibinfo {howpublished} {\url{https://reefs.com/slime-aquarists-best-friend/}} (\bibinfo {year} {2020}),\ \bibinfo {note} {reefs.com. © Reefs.com, all rights reserved. Accessed: April 2025}\BibitemShut {NoStop}%
\bibitem [{\citenamefont {Swanson}(2020)}]{Swanson2020}%
  \BibitemOpen
  \bibfield  {author} {\bibinfo {author} {\bibfnamefont {S.}~\bibnamefont {Swanson}},\ }\href@noop {} {\enquote {\bibinfo {title} {Plants sense snail slime to avoid becoming supper},}\ }\bibinfo {howpublished} {\url{https://www.discovermagazine.com/planet-earth/plants-sense-snail-slime-to-avoid-becoming-supper}} (\bibinfo {year} {2020}),\ \bibinfo {note} {discover Magazine. © University of Wisconsin–Madison. Image by Sarah Swanson. Accessed: April 2025}\BibitemShut {NoStop}%
\bibitem [{\citenamefont {Scott}(2024)}]{Scott2024}%
  \BibitemOpen
  \bibfield  {author} {\bibinfo {author} {\bibfnamefont {J.}~\bibnamefont {Scott}},\ }\href@noop {} {\enquote {\bibinfo {title} {Photo of idaho giant salamander (dicamptodon aterrimus)},}\ }\bibinfo {howpublished} {\url{https://www.inaturalist.org/photos/426241885}} (\bibinfo {year} {2024}),\ \bibinfo {note} {iNaturalist. © Jake Scott, all rights reserved. Accessed: April 2025}\BibitemShut {NoStop}%
\bibitem [{\citenamefont {Shetlar}(2024)}]{OSUExtension2024}%
  \BibitemOpen
  \bibfield  {author} {\bibinfo {author} {\bibfnamefont {D.}~\bibnamefont {Shetlar}},\ }\href@noop {} {\enquote {\bibinfo {title} {Slugs and their management in landscapes},}\ }\bibinfo {howpublished} {\url{https://ohioline.osu.edu/factsheet/hyg-2010}} (\bibinfo {year} {2024}),\ \bibinfo {note} {© The Ohio State University Extension}\BibitemShut {NoStop}%
\bibitem [{\citenamefont {Ghosh}, \citenamefont {Ebrahimi},\ and\ \citenamefont {Vaziri}(2017)}]{ghosh2017non}%
  \BibitemOpen
  \bibfield  {author} {\bibinfo {author} {\bibfnamefont {R.}~\bibnamefont {Ghosh}}, \bibinfo {author} {\bibfnamefont {H.}~\bibnamefont {Ebrahimi}}, \ and\ \bibinfo {author} {\bibfnamefont {A.}~\bibnamefont {Vaziri}},\ }\bibfield  {title} {\enquote {\bibinfo {title} {Non-ideal effects in bending response of soft substrates covered with biomimetic scales},}\ }\href@noop {} {\bibfield  {journal} {\bibinfo  {journal} {Journal of the mechanical behavior of biomedical materials}\ }\textbf {\bibinfo {volume} {72}},\ \bibinfo {pages} {1--5} (\bibinfo {year} {2017})}\BibitemShut {NoStop}%
\bibitem [{\citenamefont {Ali}, \citenamefont {Ebrahimi},\ and\ \citenamefont {Ghosh}(2019{\natexlab{a}})}]{ali2019bending}%
  \BibitemOpen
  \bibfield  {author} {\bibinfo {author} {\bibfnamefont {H.}~\bibnamefont {Ali}}, \bibinfo {author} {\bibfnamefont {H.}~\bibnamefont {Ebrahimi}}, \ and\ \bibinfo {author} {\bibfnamefont {R.}~\bibnamefont {Ghosh}},\ }\bibfield  {title} {\enquote {\bibinfo {title} {Bending of biomimetic scale covered beams under discrete non-periodic engagement},}\ }\href@noop {} {\bibfield  {journal} {\bibinfo  {journal} {International Journal of Solids and Structures}\ }\textbf {\bibinfo {volume} {166}},\ \bibinfo {pages} {22--31} (\bibinfo {year} {2019}{\natexlab{a}})}\BibitemShut {NoStop}%
\bibitem [{\citenamefont {Ali}, \citenamefont {Ebrahimi},\ and\ \citenamefont {Ghosh}(2019{\natexlab{b}})}]{ali2019frictional}%
  \BibitemOpen
  \bibfield  {author} {\bibinfo {author} {\bibfnamefont {H.}~\bibnamefont {Ali}}, \bibinfo {author} {\bibfnamefont {H.}~\bibnamefont {Ebrahimi}}, \ and\ \bibinfo {author} {\bibfnamefont {R.}~\bibnamefont {Ghosh}},\ }\bibfield  {title} {\enquote {\bibinfo {title} {Frictional damping from biomimetic scales},}\ }\href@noop {} {\bibfield  {journal} {\bibinfo  {journal} {Scientific Reports}\ }\textbf {\bibinfo {volume} {9}},\ \bibinfo {pages} {14628} (\bibinfo {year} {2019}{\natexlab{b}})}\BibitemShut {NoStop}%
\bibitem [{\citenamefont {Ebrahimi}\ \emph {et~al.}(2023)\citenamefont {Ebrahimi}, \citenamefont {Krsmanovic}, \citenamefont {Ali},\ and\ \citenamefont {Ghosh}}]{ebrahimi2023material}%
  \BibitemOpen
  \bibfield  {author} {\bibinfo {author} {\bibfnamefont {H.}~\bibnamefont {Ebrahimi}}, \bibinfo {author} {\bibfnamefont {M.}~\bibnamefont {Krsmanovic}}, \bibinfo {author} {\bibfnamefont {H.}~\bibnamefont {Ali}}, \ and\ \bibinfo {author} {\bibfnamefont {R.}~\bibnamefont {Ghosh}},\ }\bibfield  {title} {\enquote {\bibinfo {title} {Material-geometry interplay in damping of biomimetic scale beams},}\ }\href@noop {} {\bibfield  {journal} {\bibinfo  {journal} {Applied Physics Letters}\ }\textbf {\bibinfo {volume} {123}} (\bibinfo {year} {2023})}\BibitemShut {NoStop}%
\bibitem [{\citenamefont {Sarkar}\ \emph {et~al.}(2025)\citenamefont {Sarkar}, \citenamefont {Ebrahimi}, \citenamefont {Hossain}, \citenamefont {Ali},\ and\ \citenamefont {Ghosh}}]{sarkar2025bending}%
  \BibitemOpen
  \bibfield  {author} {\bibinfo {author} {\bibfnamefont {P.~R.}\ \bibnamefont {Sarkar}}, \bibinfo {author} {\bibfnamefont {H.}~\bibnamefont {Ebrahimi}}, \bibinfo {author} {\bibfnamefont {M.~S.}\ \bibnamefont {Hossain}}, \bibinfo {author} {\bibfnamefont {H.}~\bibnamefont {Ali}}, \ and\ \bibinfo {author} {\bibfnamefont {R.}~\bibnamefont {Ghosh}},\ }\bibfield  {title} {\enquote {\bibinfo {title} {Bending mechanics of biomimetic scale plates},}\ }\href@noop {} {\bibfield  {journal} {\bibinfo  {journal} {European Journal of Mechanics-A/Solids}\ ,\ \bibinfo {pages} {105664}} (\bibinfo {year} {2025})}\BibitemShut {NoStop}%
\bibitem [{\citenamefont {Ali}, \citenamefont {Ebrahimi},\ and\ \citenamefont {Ghosh}(2019{\natexlab{c}})}]{ali2019tailorable}%
  \BibitemOpen
  \bibfield  {author} {\bibinfo {author} {\bibfnamefont {H.}~\bibnamefont {Ali}}, \bibinfo {author} {\bibfnamefont {H.}~\bibnamefont {Ebrahimi}}, \ and\ \bibinfo {author} {\bibfnamefont {R.}~\bibnamefont {Ghosh}},\ }\bibfield  {title} {\enquote {\bibinfo {title} {Tailorable elasticity of cantilever using spatio-angular functionally graded biomimetic scales},}\ }\href@noop {} {\bibfield  {journal} {\bibinfo  {journal} {Mechanics of Soft Materials}\ }\textbf {\bibinfo {volume} {1}},\ \bibinfo {pages} {1--12} (\bibinfo {year} {2019}{\natexlab{c}})}\BibitemShut {NoStop}%
\bibitem [{\citenamefont {Chong}\ \emph {et~al.}(2019)\citenamefont {Chong}, \citenamefont {Hamdan}, \citenamefont {Wong},\ and\ \citenamefont {Yusup}}]{chong2019modelling}%
  \BibitemOpen
  \bibfield  {author} {\bibinfo {author} {\bibfnamefont {W.~W.~F.}\ \bibnamefont {Chong}}, \bibinfo {author} {\bibfnamefont {S.~H.}\ \bibnamefont {Hamdan}}, \bibinfo {author} {\bibfnamefont {K.~J.}\ \bibnamefont {Wong}}, \ and\ \bibinfo {author} {\bibfnamefont {S.}~\bibnamefont {Yusup}},\ }\bibfield  {title} {\enquote {\bibinfo {title} {Modelling transitions in regimes of lubrication for rough surface contact},}\ }\href@noop {} {\bibfield  {journal} {\bibinfo  {journal} {Lubricants}\ }\textbf {\bibinfo {volume} {7}},\ \bibinfo {pages} {77} (\bibinfo {year} {2019})}\BibitemShut {NoStop}%
\bibitem [{\citenamefont {Bhushan}\ and\ \citenamefont {Ko}(2003)}]{bhushan2003introduction}%
  \BibitemOpen
  \bibfield  {author} {\bibinfo {author} {\bibfnamefont {B.}~\bibnamefont {Bhushan}}\ and\ \bibinfo {author} {\bibfnamefont {P.~L.}\ \bibnamefont {Ko}},\ }\bibfield  {title} {\enquote {\bibinfo {title} {Introduction to tribology},}\ }\href@noop {} {\bibfield  {journal} {\bibinfo  {journal} {Appl. Mech. Rev.}\ }\textbf {\bibinfo {volume} {56}},\ \bibinfo {pages} {B6--B7} (\bibinfo {year} {2003})}\BibitemShut {NoStop}%
\bibitem [{\citenamefont {Dharmavaram}, \citenamefont {Ebrahimi},\ and\ \citenamefont {Ghosh}(2022)}]{dharmavaram2022coupled}%
  \BibitemOpen
  \bibfield  {author} {\bibinfo {author} {\bibfnamefont {S.}~\bibnamefont {Dharmavaram}}, \bibinfo {author} {\bibfnamefont {H.}~\bibnamefont {Ebrahimi}}, \ and\ \bibinfo {author} {\bibfnamefont {R.}~\bibnamefont {Ghosh}},\ }\bibfield  {title} {\enquote {\bibinfo {title} {Coupled bend--twist mechanics of biomimetic scale substrate},}\ }\href@noop {} {\bibfield  {journal} {\bibinfo  {journal} {Journal of the Mechanics and Physics of Solids}\ }\textbf {\bibinfo {volume} {159}},\ \bibinfo {pages} {104711} (\bibinfo {year} {2022})}\BibitemShut {NoStop}%
\bibitem [{\citenamefont {Tavakol}\ \emph {et~al.}(2017)\citenamefont {Tavakol}, \citenamefont {Froehlicher}, \citenamefont {Holmes},\ and\ \citenamefont {Stone}}]{tavakol2017extended}%
  \BibitemOpen
  \bibfield  {author} {\bibinfo {author} {\bibfnamefont {B.}~\bibnamefont {Tavakol}}, \bibinfo {author} {\bibfnamefont {G.}~\bibnamefont {Froehlicher}}, \bibinfo {author} {\bibfnamefont {D.~P.}\ \bibnamefont {Holmes}}, \ and\ \bibinfo {author} {\bibfnamefont {H.~A.}\ \bibnamefont {Stone}},\ }\bibfield  {title} {\enquote {\bibinfo {title} {Extended lubrication theory: improved estimates of flow in channels with variable geometry},}\ }\href@noop {} {\bibfield  {journal} {\bibinfo  {journal} {Proceedings of the Royal Society A: Mathematical, Physical and Engineering Sciences}\ }\textbf {\bibinfo {volume} {473}},\ \bibinfo {pages} {20170234} (\bibinfo {year} {2017})}\BibitemShut {NoStop}%
\bibitem [{\citenamefont {Hamrock}, \citenamefont {Schmid},\ and\ \citenamefont {Jacobson}(2004)}]{hamrock2004fundamentals}%
  \BibitemOpen
  \bibfield  {author} {\bibinfo {author} {\bibfnamefont {B.~J.}\ \bibnamefont {Hamrock}}, \bibinfo {author} {\bibfnamefont {S.~R.}\ \bibnamefont {Schmid}}, \ and\ \bibinfo {author} {\bibfnamefont {B.~O.}\ \bibnamefont {Jacobson}},\ }\href@noop {} {\emph {\bibinfo {title} {Fundamentals of fluid film lubrication}}}\ (\bibinfo  {publisher} {CRC press},\ \bibinfo {year} {2004})\BibitemShut {NoStop}%
\bibitem [{\citenamefont {Kholy}(2007)}]{elkholy2007granular}%
  \BibitemOpen
  \bibfield  {author} {\bibinfo {author} {\bibfnamefont {K.~N.~E.}\ \bibnamefont {Kholy}},\ }\emph {\bibinfo {title} {Granular Contact Lubrication: Theory and Experiment}},\ \href {https://repository.lsu.edu/gradschool_dissertations/1718} {Ph.D. thesis},\ \bibinfo  {school} {Louisiana State University} (\bibinfo {year} {2007})\BibitemShut {NoStop}%
\bibitem [{\citenamefont {Bari{\'c}}\ and\ \citenamefont {Steiner}(2016)}]{baric2016extended}%
  \BibitemOpen
  \bibfield  {author} {\bibinfo {author} {\bibfnamefont {E.}~\bibnamefont {Bari{\'c}}}\ and\ \bibinfo {author} {\bibfnamefont {H.}~\bibnamefont {Steiner}},\ }\bibfield  {title} {\enquote {\bibinfo {title} {Extended lubrication theory for generalized couette flow through converging gaps},}\ }\href@noop {} {\bibfield  {journal} {\bibinfo  {journal} {International Journal of Heat and Mass Transfer}\ }\textbf {\bibinfo {volume} {99}},\ \bibinfo {pages} {149--158} (\bibinfo {year} {2016})}\BibitemShut {NoStop}%
\bibitem [{\citenamefont {Coyne}\ and\ \citenamefont {Elrod~Jr}(1970)}]{coyne1970conditions}%
  \BibitemOpen
  \bibfield  {author} {\bibinfo {author} {\bibfnamefont {J.}~\bibnamefont {Coyne}}\ and\ \bibinfo {author} {\bibfnamefont {H.}~\bibnamefont {Elrod~Jr}},\ }\bibfield  {title} {\enquote {\bibinfo {title} {Conditions for the rupture of a lubricating film. part i: theoretical model},}\ }\href@noop {} {\bibfield  {journal} {\bibinfo  {journal} {Journal of Tribology}\ } (\bibinfo {year} {1970})}\BibitemShut {NoStop}%
\bibitem [{\citenamefont {Ota}(1987)}]{ota1987note}%
  \BibitemOpen
  \bibfield  {author} {\bibinfo {author} {\bibfnamefont {T.}~\bibnamefont {Ota}},\ }\bibfield  {title} {\enquote {\bibinfo {title} {A note on film rupture in hydrodynamic lubrication},}\ }\href@noop {} {\bibfield  {journal} {\bibinfo  {journal} {Journal of Tribology}\ } (\bibinfo {year} {1987})}\BibitemShut {NoStop}%
\bibitem [{\citenamefont {Nosov}\ and\ \citenamefont {Gomez-Mancilla}(2004)}]{nosov2004appearance}%
  \BibitemOpen
  \bibfield  {author} {\bibinfo {author} {\bibfnamefont {V.~R.}\ \bibnamefont {Nosov}}\ and\ \bibinfo {author} {\bibfnamefont {J.}~\bibnamefont {Gomez-Mancilla}},\ }\bibfield  {title} {\enquote {\bibinfo {title} {On the appearance of lubricant film rupture in cylindrical journal bearings},}\ }\href@noop {} {\bibfield  {journal} {\bibinfo  {journal} {Tribology transactions}\ }\textbf {\bibinfo {volume} {47}},\ \bibinfo {pages} {233--238} (\bibinfo {year} {2004})}\BibitemShut {NoStop}%
\bibitem [{\citenamefont {Alzghoul}, \citenamefont {Cabezas},\ and\ \citenamefont {Szil{\'a}gyi}(2022)}]{alzghoul2022dynamic}%
  \BibitemOpen
  \bibfield  {author} {\bibinfo {author} {\bibfnamefont {M.}~\bibnamefont {Alzghoul}}, \bibinfo {author} {\bibfnamefont {S.}~\bibnamefont {Cabezas}}, \ and\ \bibinfo {author} {\bibfnamefont {A.}~\bibnamefont {Szil{\'a}gyi}},\ }\bibfield  {title} {\enquote {\bibinfo {title} {Dynamic modeling of a simply supported beam with an overhang mass},}\ }\href@noop {} {\bibfield  {journal} {\bibinfo  {journal} {Pollack Periodica}\ } (\bibinfo {year} {2022})}\BibitemShut {NoStop}%
\bibitem [{\citenamefont {Blevins}(2015)}]{blevins2015formulas}%
  \BibitemOpen
  \bibfield  {author} {\bibinfo {author} {\bibfnamefont {R.~D.}\ \bibnamefont {Blevins}},\ }\href@noop {} {\emph {\bibinfo {title} {Formulas for dynamics, acoustics and vibration}}}\ (\bibinfo  {publisher} {John Wiley \& Sons},\ \bibinfo {year} {2015})\BibitemShut {NoStop}%
\bibitem [{\citenamefont {Christensen}(2013)}]{christensen2013theory}%
  \BibitemOpen
  \bibfield  {author} {\bibinfo {author} {\bibfnamefont {R.~M.}\ \bibnamefont {Christensen}},\ }\href@noop {} {\emph {\bibinfo {title} {Theory of viscoelasticity}}}\ (\bibinfo  {publisher} {Courier Corporation},\ \bibinfo {year} {2013})\BibitemShut {NoStop}%
\end{thebibliography}%

\end{document}